\pdfoutput=1
\documentclass[opre,nonblindrev]{informs2} % for review, not blinded
\usepackage{endnotes}
\let\footnote=\endnote
\let\enotesize=\normalsize
\def\notesname{Endnotes}%
\def\makeenmark{$^{\theenmark}$}
\def\enoteformat{\rightskip0pt\leftskip0pt\parindent=1.75em
  \leavevmode\llap{\theenmark.\enskip}}

\usepackage{natbib}
 \bibpunct[, ]{(}{)}{,}{a}{}{,}%
 \def\bibfont{\small}%
 \def\bibsep{\smallskipamount}%
 \def\bibhang{24pt}%
 \def\newblock{\ }%
 \def\BIBand{and}%

\TheoremsNumberedThrough     % Preferred (Theorem 1, Lemma 1, Theorem 2)
\ECRepeatTheorems

\EquationsNumberedThrough    % Default: (1), (2), ...
\usepackage{amsmath,amstext,amsfonts,amssymb,mathrsfs,euscript}%amsthm,
\usepackage{graphicx,color,fullpage,setspace}
\usepackage{personal}

\begin{document}

\TITLE{Strategic Arrivals into Queueing Networks:\\ The Network Concert Queueing Game}

\ARTICLEAUTHORS{%
\AUTHOR{Harsha Honnappa}
\AFF{EE Department, University of Southern California, Los Angeles, CA 90089. Email: \EMAIL{\tt honnappa@usc.edu}} %, \URL{}}
\AUTHOR{Rahul Jain}
\AFF{EE \& ISE Departments, University of Southern California, Los Angeles, CA 90089. Email: \EMAIL{\tt rahul.jain@usc.edu}}
} % end of the block

%\KEYWORDS{deterministic inventory theory; infinite linear programming duality;
%  existence of optimal policies; semi-Markov decision process; cyclic schedule}


%\vspace*{-33pt}\begin{center}\date{\today} \\

\ABSTRACT{
Queueing networks are typically modelled assuming that the arrival process is exogenous, and unaffected by admission control, scheduling policies, etc.  In many situations, however, users choose the time of their arrival strategically, taking delay and other metrics into account.  In this paper, we develop a framework to study such strategic arrivals into queueing networks. We start by deriving a functional strong law of large numbers (FSLLN) approximation to the queueing network. In the fluid limit derived, we then study the population game wherein users strategically choose when to arrive, and upon arrival which of the $K$ queues to join. The queues  start service at given times, which can potentially be different. We characterize the (strategic) arrival process at each of the queues, and the price of anarchy of the ensuing strategic arrival game. We then extend the analysis to multiple populations of users, each with a different cost metric. The equilibrium arrival profile and price of anarchy are derived. Finally, we present the methodology for exact equilibrium analysis. This, however, is tractable for only some simple cases such as two users arriving at a two node queueing network, which we then present. 
}
%\HISTORY{Submitted: December 12, 2011}

\maketitle

%===============================================================================

\section{Introduction} \label{sec:intro}
This paper is motivated by the following scenario: Users arriving at a concert, a game or at a store for Black Friday sales, where arriving before others is preferable, are faced with the dilemma of when to arrive. Should one arrive early before others and wait a while for service to start, or arrive late and wait less, and yet by which time the best seats or deals may already be gone? In such settings, when rational users make strategic decisions of timing, we cannot assume that the arrival process can be modelled by an exogenous renewal process such as a Poisson process. Furthermore, there may be multiple queues (which may start service at different times) and arriving users may have a choice of which queue to join.  
%For example, airport passengers must choose when to arrive at the airport, and upon arrival which security queue to join. 

Similarly, users downloading large files from a website often time their downloads to times of day when network congestion is expected to be lower (e.g., late at night.) Moreover, upon arrival (at the web-site), they may have to choose which server to download from. A natural question to ask is ``Does an equilibrium arrival process exist"? If it does, is it efficient with respect to some metric? If not, can we bound the amount of inefficiency? We answer these questions by modeling this strategic arrival behavior as a game that we call the \emph{network concert queueing game}. Such strategic analysis of  queues was introduced in \cite{JuJa2009,JaJuSh2010a} for a single server FIFO queue. In this paper, we extend that analysis to a network of queues.

%For example, airport passengers on a flight arrive at the security check point in a single stream. The passengers are apriori aware that there will be $N$ parallel queues to pass through and also know the service rate at each queue. Every passenger aims to minimize the amount of time spent waiting for service, while also ensuring that the time of arrival is close enough to the departure time. That is, passengers will choose a time of arrival and a particular queue to join \emph{strategically}. This results in a game. A natural question to ask is whether an equilibrium exists in this game. If an equilibrium exists, what is the arrival time distribution of passengers? Is it unique? What is the efficiency of this equilibrium in relation to a socially optimal outcome?

In this paper, users choose their arrival time into a parallel queueing network wherein queues serve at different rates, and start service at different times. We assume servers are work-conserving with infinite buffers. Users can start to queue even before service starts, and do not renege or balk. We also consider that users may belong to multiple populations, each with different cost characteristics. The game is analyzed in the fluid approximation setting which offers significant analytical simplicity and tractability, while still capturing essential features of the problem. Each arriving user chooses a queue to join, and a time to arrive that minimizes a linear cost that is a weighted function of the waiting time and the service completion time. The service completion time of a user depends on the arrival time, and may be considered as a proxy for the latter metric.

We make two main contributions in this paper. Our first contribution is a fluid limit for the queueing model we introduce, wherein each user picks his time of arrival from a distribution. Thus, the inter-arrival times are not independent, and the arrival process need not be a renewal process in general. Functional strong law of large numbers approximation to the queue length process, the busy time and the virtual waiting time process at each queue is derived. Also of interest would be a diffusion limit (via the functional central limit theorem) for the various processes of interest in this non-standard queueing model. We have made some progress on this, and is on-going work.

Our second contribution is to take the fluid approximation derived above, and study the associated population game wherein each non-atomic user strategically picks a time of arrival, as well as a queue it would join. Existence of equilibria in such non-atomic games was established in \cite{Sc1973}.  We, however, argue its existence (and uniqueness) by construction. The equilibrium arrival profile turns out to be a uniform distribution over a time interval that we can determine. We also characterize the loss in social welfare due to strategically arriving users, and obtain an exact expression for the price of anarchy of the game. 

%The contributions of this paper over prior work include an explicit derivation of the fluid approximation to the queueing network by developing path-wise approximations to the network parameters such as the queue length, busy time and workload processes. These are then used to analyze strategic arrivals of users from a homogeneous population into a network of $K$ parallel queues. Equilibrium arrival distributions and processes are established. An exact expression for price of anarchy in terms of exogenous parameters is also computed, and it is shown that it is minimized when all the queues start operation at the same time. We characterize the equilibrium arrival profile, and derive the price of anarchy, in the case of $N$ distinct populations (each with different cost characteristics) arriving at the network. 

%we extended these results to the case of a two queue parallel network with a single arriving population and two populations, with differing delay risk characteristic. We showed that the arrivals to each of the two queues is still uniform over a compact set interval. We also computed the price of anarchy of the game and showed that it is at least 2.

While there has been a lot of work on studying pricing of queueing service (see \cite{DuJa11, HaHa2003}, \cite{MeWh1990, Na1969}), games of timing where the users choose an arrival time strategically are not so well-understood. The earliest such work is \cite{GlHa1983} in which a discrete population of users choose the time of their arrival strategically into an ?/M/1 queue, by minimizing the queueing delay. Problems with similar motivation have been considered in the transportation literature but they have focused on non-queueing theoretic fluid models with delay alone as a cost metric \citep{Li2004}. In contrast, the framework of \cite{JaJuSh2010a} is more general: Each user has a cost which is a function of the waiting time as well as the service completion time of the user (a proxy for the number of users who arrive before that user) - a significant motivation for users to arrive early in many scenarios. On the other hand, only a fluid approximation of the discrete population model is considered.  
%In this paper, we extend this setting to the case when users have a choice of being serviced by any of the multiple queueing service providers. 

% We should note here that the sense of multiple populations is not
% similar to the notion of multi-class queueing networks. Here, the
% arrivals to the network from a particular population share the same
% delay risk and have homogeneous service time
% requirements. Disparate populations have their own delay risk, but could
% share the same service requirements. Thus, the arrivals to the
% parallel queue network actually form a single stream of homogeneous
% users. Thus, this queueing network can be considered a generalized
% Jackson network.\\

In Section \ref{sec:fluid-limit}, we first develop a path-wise description of the parameters describing the queueing network, and fluid limit approximations to these processes. Next, we analyze the strategic arrivals game in the fluid setting, for a single arriving population in Section \ref{sec:strategic-arrivals}, and derive the equilibrium arrival distribution and the price of anarchy of this game. In Section \ref{sec:multiple}, we derive the equilibrium arrival profile for multiple populations with disparate arrival costs, and show that it is unique. We derive the price of anarchy of this game, show that in a special case, it is bounded above by 2. In Section \ref{sec:finite},  we illustrate the difficulty in doing exact equilibrium analysis for a finite population, and hence the importance of equilibrium analysis in the fluid limit. We conclude with a summary and discussion of further work in Section \ref{sec:conclusion}.

%----------------------------------------------------------------------------------------------------
\section{The Queueing Network: Model and Fluid Limit} \label{sec:fluid-limit}

Consider a queueing network with $K$ single server FIFO nodes in parallel. Each node starts service at some fixed time (which could be different from the other nodes), offers service with a finite service rate, and operates independently of the other nodes. Each users' \emph{time of  arrival} is an i.i.d. random variable with a given distribution and known support such that users can arrive and queue up even before service starts. Note that inter-arrival times need not be independent. Service times are i.i.d., and independent of the arrival process.  
%A useful way to interpret this process is as a case of an arrival process with general inter-arrival time distribution. 
A user upon arrival joins one of the queues according to a given routing probability. Routing will be assumed independent of the arrival time process. 
%Thus, the queueing network under consideration is a generalized (parallel queue) Jackson network.%  depicted in Figure \ref{fig:single}
We first develop the fluid limits for the queueing model introduced as the population size increases to $\infty$.
% We designate this network by $\mathcal{G} = (K,p)$. x
% In the ensuing discussion we will first give path-wise descriptions of the network parameters and then use these to develop the functional strong law of large numbers approximations to these processes, as the number of arrivals increases to $\infty$.

%Note that we use the terms \emph{fluid limit} and \emph{functional strong law of large numbers limit} interchangeably.\\

Let $(\Omega, \mathcal{F},P)$ be a probability space, and $\mathcal{D}^K := \mathcal{D}^K[-T_0,\infty)$ the space of $K$-dimensional ``right-continuous with left limits'', or \emph{cadlag} processes (see \cite{Du2010,Bi1968}). Suppose there are $n$ users arriving at the queueing network. Let $\xi_i : \Omega \rightarrow \{1,\cdots,K\}$ be a random variable such that user $i$ is routed to the $\xi_i(\omega)$th node in the network. Thus, $p_{0,k} = P(\xi_i = k)$.

%We start by describing the aggregate arrival process that counts the number of arrivals up to time $t$. 
Let $T_i : \Omega \rightarrow [-T_0,T]$ be the arrival time of user $i$, and $F$ the arrival time distribution. Denote $\mathbf{A}^n(t) := (A^n_1(t),\cdots,A^n_K(t)) \in \mathcal{D}^K$, where $A^n_K(t)$ counts the number of arrivals at queue $k$, i.e.,
\begin{eqnarray} \label{arrival-fluid-scale}
A^n_k(t) := \sum_{i=1}^n \mathbf{1}_{\{T_i \leq  t\}} \mathbf{1}_{\{\xi_i = k\}}, \quad \forall t \in [-T_0,T],
\end{eqnarray}
where $\mathbf{1}_{\{\cdot\}}$ denotes the indicator function. Let $A^n(t) = \sum_{i=1}^K A^n_i(t)$ be the aggregate number of arrivals to the network by $t$. 

Let $\nu_i^k : \Omega \rightarrow (0,\infty)$ be the service time of arrival $i$ to queue $k$ in the network, independent of $T_i$ and $\xi_i$. Let $m_k$ be the mean service time, $\mu_k = 1/m_k$ the service rate and $T_{s,k} \geq 0$ the service start time of queue $k$. % Then, the number of potential service completions by time $t$ at node $k$ is given by 
% \(
% S_k(t) := \sup \{m \in \mathcal{N} : \sum_{i=1}^m \nu_i^k \leq t-T_{s,k} \},  \quad \forall t \geq T_{s,k}.
% \)
% The vector service process is $\mathbf{S}(t) = (S_1(t), \cdots, S_K(t))^{'} \in \mathcal{D}^K$.
We \emph{accelerate} the service rate by scaling the service time by $n$, i.e.,  $\bar{\nu}_i^k = \frac{\nu_i^k}{n}$ is the accelerated service time of user $i$ at queue $k$. The accelerated or \textit{scaled service process} of queue $k$ is defined as
\begin{equation} \label{service-fluid-scale}
S_k^n(t) := \sup \bigg \{m \in \mathcal{N} \,|\, \sum_{i=1}^m \bar{\nu}_i^k \leq t - T_{s,k} \bigg \}, \quad \forall t \geq T_{s,k}.
\end{equation}
The service vector process is
$\mathbf{S}^n(t) := (S^n_1(t), \cdots, S^n_K(t))^{'} \in \mathcal{D}^K$.

Let $V_k(m) = \sum_{i=1}^m \nu^k_i$ be the service time requirement for $m$ users at queue $k$. We will call it the \textit{(cumulative) workload process}, and its \textit{scaled} process description is
\begin{equation} \label{service-time-fluid-scale}
V_k^n(t) := \sum_{i=1}^{\lfloor n t \rfloor} \bar{\nu}_i^k, \quad \forall t \in [-T_0,T].
\end{equation}
The vector cumulative workload process is $\mathbf{V}^n(t) := (V_1^n(t), \cdots, V_K^n(t))^{'}$. It is easy to see that $V_k^n(t)$ and $S_k^n(t)$ are renewal ``inverses'' of each other.

We now develop path-wise functional strong law of large numbers approximations to these processes as $n \to \infty$. We will assume that $\xi_i, T_i \text{ and } \nu_i^k$ are mutually independent. Denote $\mathbf{M} = \text{diag}(m_1,\cdots,m_K)$,  $\mathbf{p} = (p_{0,1},\cdots, p_{0,K})^{'}$, and $\mathbf{t_{s,+}}(t) := ((t-T_{s,1})_+, \cdots, (t-T_{s,K})_+)^{'} $ where $(x)_+:=x\mathbf{1}_{\{x \geq 0\}}$. The fluid limits for the arrival, service and the workload processes are then given by:
\begin{theorem} \label{thm:fluid-primitives}
As $n \to \infty$, \(
\left (\frac{1}{n} \mathbf{A}^n(t), \frac{1}{n} \mathbf{S}^n(t),
\mathbf{V}^n(t+T_0) \right ) \rightarrow \left ( \mathbf{p}F(t), \mathbf{M}^{-1}
\mathbf{t_{s,+}}(t), \mathbf{M} \mathbf{1}(t+T_0) \right ) \) 
a.s. \quad u.o.c., ~~$\forall t \in [-T_0,\infty)$, 
where $\mathbf{1} = (1,\cdots,1)^{'}$.
\end{theorem}
Here, and in the ensuing, \textit{u.o.c.} denotes ``converges uniformly on compact sets''. %The proof of the above theorem can be found in the appendix.
The proof of this theorem is omitted, as it uses standard arguments: Convergence of the service and workload processes are standard, and follow from the functional strong law of large numbers. The convergence of the arrival process follows from a generalization of the Glivenko-Cantelli Theorem, using independence of the routing and arrival time random variables. %Since each node operates independently of the others, the vector result follows trivially. 

Now, let $\mathbf{Q}(t) := (Q_1(t), \cdots, Q_K(t))^{'} \,\, \in \mathcal{D}^K$, where $Q_k(t)$ is the queue length at node $k$ at time $t$. We assume that the queueing network starts empty, so $\mathbf{Q}(-T_0) = \mathbf{0}$. The queue length process of node $k$ is given by $Q_k(t) = (A_k(t) - S_k(t))_+$, the non-negative difference between the aggregate number of arrivals and the (potential) service process up to time $t$. Let the amount of time in $[T_{s,k},t]$ that node $k$ spends serving users be called the \textit{busy time process},
\(
B_k(t) = \left (\int_{T_{s,k}}^t \mathbf{1}_{\{Q_k(s) > 0\}} ds \right ) \mathbf{1}_{\{t \geq T_{s,k}\}}.
\)
It is easy to see that $S_k(B_k(t))$ is the number of arrivals served by time $t$.
%\begin{eqnarray} Q_k(t) := A_k(t) - S_k(B_k(t)).\end{eqnarray}
Let the scaled \textit{queue length vector process} be given by
\(
Q_k^n(t)/n:= (A_k^n(t)/n - S^n_k(B^n_k(t)))_+/n.
%\frac{1}{n}\mathbf{Q}^n(t) = \frac{1}{n} \mathbf{A}^n(t) -
%\frac{1}{n} \mathbf{S}^n(B^n(t)),
\)
%We proceed by considering the $k$th component of this vector, %check the joint convegence here
%\begin{eqnarray*}
%\nonumber
%\end{eqnarray*}
%Adding and subtracting $p_{0,k}F(t)$, $\mu_k B^n_k(t)$ and $\mu_k t \mathbf{1}_{\{t \geq T_{s,k}\}}$ and $\tilde{I}_k^n(t) := \int_{-T_0}^t \mathbf{1}_{\{Q_k^n(s) = 0\}} ds$ to the right hand side of the equation above we have
We can rewrite this in vector form as
%\begin{eqnarray}\label{queue-length-path}
%\nonumber
\(
\frac{1}{n} \mathbf{Q}^n(t) = \left ( \frac{1}{n} \mathbf{A}^n(t) -
\mathbf{p} F(t) \right ) - \left ( \frac{1}{n} \mathbf{S}^n(B^n(t)) - \mathbf{M}^{-1} \mathbf{B}^n(t) \right ) + \left ( \mathbf{p} F(t) - \mathbf{M}^{-1} \mathbf{t_{s,+}}(t) \right ) 
%\label{vector-queue-length-fluid-scale}
+ \mathbf{M}^{-1} \left (\mathbf{I}^n(t) -
\tilde{\mathbf{I}}^n(t) \right ) + \mathbf{M}^{-1} \tilde{\mathbf{I}}^n(t),
\)
% \frac{1}{n} \mathbf{Q}^n(t) = \bigg ( \frac{1}{n} \mathbf{A}^n(t) -
% \mathbf{p} F(t) \bigg ) - \bigg ( \frac{1}{n} \mathbf{S}^n(B^n(t)) - \mathbf{M}^{-1} \mathbf{B}^n(t)\bigg ) + [\mathbf{p} F(t) -
% \mathbf{M}^{-1} \mathbf{t_{s,+}}(t)] 
% %\label{vector-queue-length-fluid-scale}
% + \mathbf{M}^{-1} [\mathbf{I}^n(t) -
% \tilde{\mathbf{I}}^n(t)] + \mathbf{M}^{-1} \tilde{\mathbf{I}}^n(t),
%\end{eqnarray}
where $\mathbf{B}^n(t) := (B^n_1(t), \cdots, B^n_K(t))^{'}$ is the busy time process vector, $\mathbf{I}^n(t) := (I^n_1(t), \cdots, I^n_K(t))^{'} = \mathbf{t_{s,+}}(t) - \mathbf{B}^n(t)$ is the \textit{idle-time process vector}, and $\tilde{\mathbf{I}}^n(t) := (\tilde{I}^n_1(t), \cdots, \tilde{I}^n_K(t))^{'}$ is the \textit{cumulative idleness process vector}, with $\tilde{I}_k^n(t) := \int_{-T_0}^t \mathbf{1}_{\{Q_k^n(s) = 0\}} ds$.
%\begin{eqnarray*}
%\frac{Q_k^n(t)}{n} &=& \bigg (\frac{1}{n} A_k^n(t) - p_{0,k} F(t) \bigg ) - \bigg ( \frac{1}{n} S^n_k(B^n_k(t)) - \mu_k B^n_k(t) \bigg )\\ &&+ [ p_{0,k} F(t) - \mu_k t \mathbf{1}_{\{t \geq T_{s,k}\}} ] + \mu_k [I_k^n(t) - \tilde{I}_k^n(t)] + \mu_k \tilde{I}_k^n(t), \end{eqnarray*}
%where the idle time process for the $k$th node is $I_k^n(t) = t \mathbf{1}_{\{t \geq T_{s,k}\}} - B_k^n(t) = (\int_{T_{s,k}}^t \mathbf{1}_{\{Q_k^n(s) = 0\}} ds) \mathbf{1}_{\{t \geq T_{s,k}\}}$, which measures the amount of time in $[T_{s,k},t]$ that the node is idle (i.e., the queue is empty). %in order to satisfy the sufficient conditions of the Oblique Reflection Mapping Theorem; see Theorem Theorem 7.2, \cite{ChYa2001} and Lemma \ref{lem:idle} in the Appendix. 

This can be rewritten again as
\(
\frac{1}{n} \mathbf{Q}^n(t) = \mathbf{X}^n(t) + \mathbf{Y}^n(t),
\)
where
\begin{eqnarray}
\label{X-fluid-scale}
\mathbf{X}^n(t) &:=& \bigg ( \frac{1}{n} \mathbf{A}^n(t) -
\mathbf{p} F(t) \bigg ) - \bigg ( \frac{1}{n} \mathbf{S}^n(B^n(t))
- \mathbf{M}^{-1} \mathbf{B}^n(t)\bigg ) + [\mathbf{p} F(t) -
\mathbf{M}^{-1} \mathbf{t_{s,+}}(t)] \nonumber\\
&&+ \mathbf{M}^{-1} [\mathbf{I}^n(t) -
\tilde{\mathbf{I}}^n(t)] \quad \text{ and } \quad \\
\label{Y-fluid-scale}
\mathbf{Y}^n(t) &:=& \mathbf{M}^{-1} \tilde{\mathbf{I}}^n(t),
\end{eqnarray}
It can be shown that the process $\mathbf{X}^n(t)$ satisfies the following functional strong law of large numbers.
\begin{proposition} \label{prop:X-fluid}
As $n \to \infty$, \(
\mathbf{X}^n(t) \rightarrow \bar{\mathbf{X}}(t) := (\mathbf{p} F(t) -
\mathbf{M}^{-1} \mathbf{t_{s,+}}(t)), \quad \text{a.s.} \quad \text{u.o.c.} \,\, \forall t \in [-T_0,\infty).
\)
\end{proposition}
The proof is in the appendix. We can now establish the fluid limit for the queue length process. 
\begin{theorem} \label{thm:queue-fluid}
\begin{itemize}
\item[(i)] The scaled stochastic process vector $ \left( \frac{1}{n}\mathbf{Q}^n(t), \mathbf{Y}^n(t) \right )$ satisfies the Oblique Reflection Mapping Theorem (see \cite{ChYa2001}) and,
\item[(ii)]  as $n \rightarrow \infty$,
\(
  \left ( \frac{1}{n} \mathbf{Q}^n(t), \mathbf{Y}^n(t) \right ) \rightarrow
  \left (\bar{\mathbf{Q}}(t),\bar{\mathbf{Y}}(t) \right ) :=  \left ( \Phi(\bar{\mathbf{X}}(t)), \Psi(\bar{\mathbf{X}}(t)) \right ) \quad \text{a.s.} \quad \text{u.o.c.} \,\, \forall t \in [-T_0,\infty),
\)
where $\Phi(\bar{\mathbf{X}}(t)) = \bar{\mathbf{X}}(t) + \Psi(\bar{\mathbf{X}}(t))$ and $\Psi(\bar{X})(t) = \sup_{-T_0 \leq s \leq t} (-\bar{\mathbf{X}}(s))^+$ is the reflection regulator map.
\end{itemize}
\end{theorem}
% The proof is relegated to the Appendix. Note that the regulator map is the unique fixed point, $y^* \in \mathcal{D}^K$, of the map $\pi(x,y)(t) := \sup_{-T_0 \leq s \leq t} (-x(s) + G y(s))^+ \,\, \forall t \in [-T_0,\infty)$, where $G$, an $M$-matrix, is the internal routing matrix of the network; see Chapter 7 of \cite{ChYa2001} for details. For a parallel queue network, $G = \mathbf{0}$. Thus, we have $\Psi(\bar{\mathbf{X}}(t)) = \sup_{-T_0 \leq s \leq t} (-\bar{\mathbf{X}}(s))^+$; it follows that
% \(
% (\bar{\mathbf{Q}}(t), \bar{\mathbf{Y}}(t)) = \bigg ( \bar{\mathbf{X}}(t) + \sup_{-T_0 \leq s \leq t} (-\bar{\mathbf{X}}(s))^+,\, \sup_{-T_0 \leq s \leq t} (-\bar{\mathbf{X}}(s))^+ \bigg ).
% \)

%Recall that the busy time process of the $k$th queue is given by $B^n_k(t) = \bigg ( \int_{T_{s,k}}^t \mathbf{1}_{ \{ Q^n_k(s) > 0 \} } ds \bigg ) \mathbf{1}_{\{t \geq T_{s,k}\}}$. 
We now define the \textit{virtual waiting time process} as
%\begin{equation}\label{eq:virtualwaitingtime}
\(
\mathbf{W}(t) := (W_1(t), \cdots, W_K(t))^{'} = \mathbf{Z}(t) - \mathbf{t_{s,-}}(t), 
\)
%\end{equation}
where $\mathbf{t_{s,-}}(t) = ((t-T_{s,1}) \mathbf{1}_{\{t \leq T_{s,1},\}}, \cdots, (t-T_{s,K}) \mathbf{1}_{\{t \leq T_{s,K}\}})^{'}$, and the \textit{workload process} as $\mathbf{Z(t)} := (Z_1(t), \cdots,Z_K(t))^{'}$ where
\(
Z_k(t) := V_k(A_k(t)) - B_k(t), \quad \forall t \in [-T_0,\infty).
\)
Here, $V_k(A_k(t))$ is the total work presented to queue $k$ by arrivals up to time $t$. Thus, the workload process is the amount of work remaining after the queue has been busy for $B_k(t)$ units of time in $[T_{s,k},t]$. Now, the scaled $k$th component of $\mathbf{W}(t)$ is given by
\(
W_k^n(t) = V^n_k  ( A^n_k(t) ) - B^n_k(t) -(t - T_{s,k}) \mathbf{1}_{\{t \leq T_{s,k}\}}.
\)
Theorem 3 establishes fluid limits to these processes.
\begin{theorem}\label{thm:fluid-busy-virtual}
As $n \to \infty$,
\(
\mathbf{B}^n(t) \rightarrow \bar{\mathbf{B}}(t) := \mathbf{t_{s,+}}(t) - \mathbf{M}\Psi(\bar{\mathbf{X}}(t))
 = \mathbf{t_{s,+}}(t) - \mathbf{M}\sup_{-T_0 \leq s \leq t} (-\bar{\mathbf{X}}(s))^+
\)
and
\(
\mathbf{W}^n(t) \rightarrow \bar{\mathbf{W}}(t) := \mathbf{M}
\mathbf{p} F(t) - \bar{\mathbf{B}}(t) - \mathbf{t_{s,-}}(t) \quad \text{a.s.} \quad \text{u.o.c.}, \,\, \forall t \in [-T_0, \infty).
\)
\end{theorem}
The proofs can be found in the Appendix.

%-----------------------------------------------------------------------------------------------------
\section{The Network Concert Queueing Game} \label{sec:strategic-arrivals}

We next address the following question: If the arriving users into a parallel queue network choose a time of arrival so as to minimize a cost function that trades off the amount of time spent waiting for service against the service completion time (a proxy for the number of users that arrive ahead of them), what does the arrival process look like? We consider users choosing mixed strategies, i.e., probability distributions over arrival times, and look for the mixed-strategy Nash equilibrium of the non-atomic game, derived from the fluid limit in the previous section.

Suppose the population size is $n$. We consider cost functions that are a weighted-linear combination of the mean waiting time and service completion time. Thus, the expected cost seen by a user arriving at time $t$ to queue $k$ is
\(
  C_k(t) = \alpha W_k(t) + \beta t^c_{k},
\)
where $W_k(t)$ is the expected virtual waiting time, and the service completion time $t^c_{k}$ is easily seen to be $t^c_{k} = t + m_k + w_k(t)$, with $m_k$ the mean service time. We look for symmetric equilibria in the one-shot arrival game, with each user having the cost function $C_k$. In general, it is quite difficult to obtain closed form solutions to the associated fixed-point problem (we illustrate the methodology in section \ref{sec:finite}). Thus, we scale the population size $n$ to $\infty$, and study the population game associated with the fluid limit derived in the previous section. Now, the scaled cost function is $C_k^n(t) = \a w_k^n(t) + \b t^{c,n}_{k}$. From Theorem \ref{thm:fluid-busy-virtual}, we then have the limit as $C_k(t) = \a \bar{w}_k(t) + \b \bar{t}^c_{k}$, where $\bar{t}_{c,k} = t + \bar{w}_k(t)$ and $\bar{w}_k(t) = m_k (\bar{X}(t) + \Psi(\bar{X}(t)) - (t-T_{s,k})\mathbf{1}_{t \leq T_{s,k}})$, with $\bar{X}(t) = F_k(t) - \mu_k (t-T_{s,k}) \mathbf{1}_{t \geq T_{s,k}}$. Here, $F_k(t) = \sum_{j=1}^M F_{jk}(t)$ denotes the aggregate arrival profile to queue $k$.

Now, consider $N$ populations of users, with population $j$ users having cost characteristics $(\alpha_j,\beta_j)$. Denote by $T_{s,k} \geq 0$, the time at which queue $k$ starts service.  Let  $F_{jk}(t)$ denote the arrival strategy of population $j$ at queue $k$, and $F=(F_1,\cdots,F_K)$ denotes the (aggregate) \textit{arrival profile}. Denote $\textbf{F}^j = (F_{jl}, l=1,\cdots,K)$  and  $\textbf{F} = (\textbf{F}^j, j=1,\cdots,N)$ as the \textit{strategy profile}. The service completion time for a population $j$ user arriving at time $t$ at queue $k$  is given by $t^c_{j,k} = t + W_{j,k}(t)$, where $W_{j,k}(t)$ is the virtual waiting time. Thus, the cost for a population $j$ user arriving at time $t$ at queue $k$ under arrival distribution $\textbf{F}$ is
\(
  C_{j,k,\textbf{F}}(t) = \alpha_j W_{j,k}(t) + \beta_j (t + W_{j,k}(t)).
\)
We now define (symmetric) mixed strategy Nash equilibrium profiles for the non-atomic/population game. 
\begin{definition}\label{def:eqprofile}
A strategy profile $\textbf{F}$ is an {\em equilibrium strategy profile} if for each population $j$, $\textbf{F}^j$ is a minimizer of the corresponding cost functions $C_{jk}$ at each queue $k$ at every time $\tau$ in the support of $F_{jk}$ (denoted $\mathcal{T}_{jk}$), i.e., for any arrival profile  $\textbf{G}$, and $\forall j$, $\forall k$,
\[
  C_{jk,\textbf{F}}(\tau) \leq C_{jk,\textbf{G}}(t) \quad \forall \tau \in
  \mathcal{T}_j, \quad -\infty <t<\infty.
\]
\end{definition}
In recent literature, these have been called \textit{mean field equilibria} \citep{AdJo10}. 
%This corresponds to a \emph{Wardrop equilibrium} of the strategic arrivals game.

The equilibrium condition captures the fact that for each population, the equilibrium profile must minimize its cost into any queue at any time.  Furthermore, it also implies that all queues with a positive flow of population $j$ must have equal cost, i.e., $~\forall l,k$,
\(
C_{jl,\textbf{F}}(t)\mathbf{\underline{1}}\{F_{jl}(\infty)>0\} = C_{jk,\textbf{F}}(t^{'}) \mathbf{\underline{1}}\{F_{jk}(\infty)>0\},
\)
for all $t$ and $t^{'}$ in the support of (the Lebesgue measure corresponding to) $F_{jl}$ and $F_{jk}$ respectively. Though it can be established herein, due to space constraints, we will just assume that an equilibrium arrival profile does not have any singular continuous components.
%\subsection{Single Population} \label{sec:single-eq}
Without loss of generality, we also assume that server 1 starts service at $T_{s,1}=0$ while other servers have a delayed start with $T_{s,i} > 0$, $i \neq 1$. For simplicity, assume that the queues ${1, \cdots, K}$ start in that order. In the rest of this section, we will consider only a single population of users.

% The cost function for a user arriving at queue $i$ is given by $C_i(t) = (\alpha + \beta) t^c_{i }- \alpha t$, where $t^c_{i}$ is the time of service completion given by  $t^c_{i} = W_i(t) + t$.
% We denote the time of first arrival into queue $l$ by $-T_{0,l}$, the time of last arrival into any queue by $T$, and the time the last user served departs from queue $l$ by $T_{fl}$. The next two Lemmas help in finding the equilibrium arrival profile.  
We denote the time of first arrival into queue $l$ by $-T_{0,l}$, the time of last arrival into any queue by $T_l$, and the time the last user served departs from queue $l$ by $T_{fl}$. The next two Lemmas help in finding the equilibrium arrival profile.  
\begin{lemma} 
\label{lem:equilibrium-end-time}
At equilibrium, all queues finish serving users at the same time instant. 
\end{lemma}
%The proof follows that of Lemma 1 in \cite{HoJa2010}, and is included here for the reader's convenience.
\Proof
For simplicity, consider only two queues. To see that $T_{f1}=T_{f2}$, note that at equilibrium, the costs at each of the queues must be equal at all times. Assume that $T_{f1} < T_{f2}$. Then a user arriving into queue 2 at any time $ t \in (T_{f1},T_{f2}]$ must experience a higher cost than if she had simply joined queue 1 (which is now idle). It follows that this arrival profile cannot be an equilibrium. Similarly, $T_{f2} \not < T_{f1}$. Thus, we must have $T_{f1}=T_{f2}$. 
\EndProof
The following fact is intuitively obvious but we give a formal argument.
\begin{lemma}
\label{lem:idle-equilibrium}
The parallel queue network is never idle at equilibrium.
\end{lemma}
%The proof is similar to that of Lemma 3 in \cite{JaJuSh2010a}, and is included here for the reader's convenience.
\Proof
We prove this by contradiction. In light of the fact that the cost of arriving at any of the queues is the same, it suffices to prove the assertion in the case of a single queue alone. Let $T_{s} = 0$, the service start time. Let $F$ be such that $t^* := inf \{-T_0 \leq s \leq T | \Psi(\bar{X}(t)) > 0\}$, i.e., $t^*$ is the first time that the regulator mapping is positive during the arrival interval, implying the queue is idle. This implies that $\Psi(\bar{X}(t)) = -(F(t^*) - \mu t^* \mathbf{1}_{\{t^* \geq 0\}}) > 0$. Now, let $\epsilon > 0$, and $t^*-\epsilon$ is just to the left of $t^*$. Then, it follows that, $C(t^*) = \beta t^*$. Consider,
\(
C(t^*) - C(t^*-\epsilon) = \beta t^* - \frac{(\alpha + \beta)}{\mu}(\bar{Q}(t^*-\epsilon)) - \beta (t^*-\epsilon).
\)
As $t^*$ is the first time the queue is empty, it follows that $\bar{Q}(t^*-\epsilon) = (F(t^*-\epsilon) - \mu (t^*-\epsilon) \mathbf{1}_{\{t^*- \epsilon \geq 0\}})$. Substituting for $\bar{Q}$ in the expression above, we have
\(
C(t^*) - C(t^*-\epsilon) = \beta(\epsilon) - \frac{(\alpha+\beta)}{\mu} (F(t^*-\epsilon) - \mu (t^*-\epsilon) \mathbf{1}_{\{t^*-\epsilon \geq 0\}}).
\)
Let $\epsilon \rightarrow 0$, and use the fact that $F$ has no singularities (and hence is continuous), it follows that $C(t^*) < C(t^*-)$. This implies that this arrival profile $F$, cannot be an equilibrium, thus proving the claim.
\EndProof
%\noindent \textbf{Remark.} We note that the Lemmas above hold in the fluid limit only.

It follows from Lemmas \ref{lem:equilibrium-end-time} and \ref{lem:idle-equilibrium} that at equilibrium, with a homogeneous population, the last arrivals into any queue should all happen at the same instant, and this time coincides with the instant at which the service process catches up with the backlog; that is, $T_{fl} = T_l = T$. Using the above Lemmas, it follows that the cost to a (non-atomic) user arriving at time $t$ as
\begin{eqnarray}
\mathbf{C}(t) = (\alpha + \beta)(M\bar{\mathbf{X}}(t) - \mathbf{t}_{s,-}(t)) + \beta t \mathbf{\underline{1}}
% &=& 
% \begin{bmatrix}
% (\alpha + \beta) \{p_{0,1}\frac{F(t)}{\mu_1} - (t - T_{s,1})\} + \beta t\\
% \vdots\\
% (\alpha + \beta) \{p_{0,K}\frac{F(t)}{\mu_1} - (t - T_{s,K}) \} + \beta t
% \end{bmatrix}\\ 
%\end{eqnarray*}\begin{eqnarray*}
= 
\begin{bmatrix}
(\alpha + \beta) ( p_{0,1} \frac{ F(t)}{\mu_1} + T_{s,1} ) -\alpha t\\
\vdots\\
(\alpha + \beta) ( p_{0,K} \frac{F(t)}{\mu_K} + T_{s,K} ) -\alpha t
\end{bmatrix}.
\end{eqnarray}

%.....................................................................................................................
\subsection{Arrival Distribution} \label{sec:singledist}

\begin{figure}[ht] 
\centering
\includegraphics[scale=0.8]{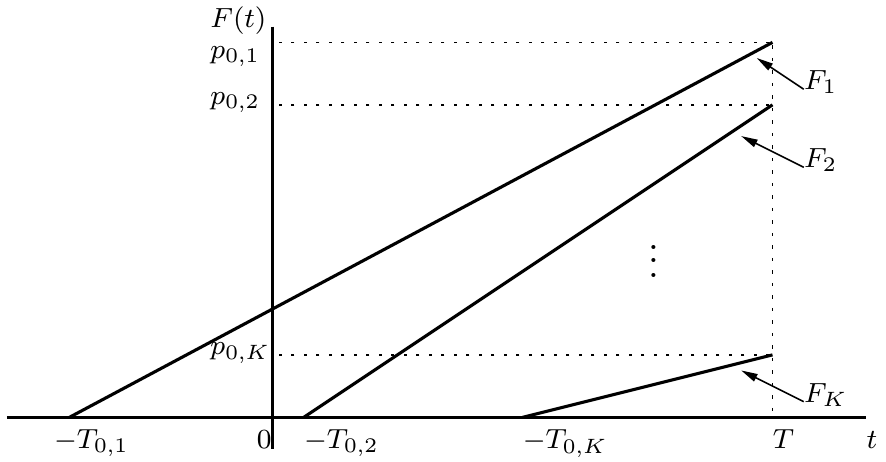}
\caption{Equilibrium arrival profile of a
  single population to a  $K$-queue parallel queueing network.}
\label{fig:singledist}
\end{figure}

Now, let the arrival profile at queue $l$ at equilibrium be $F_l^*(t)$ with support on $[-T_{0,l},T]$, and define $F_l^*(t):=p_{0,l} F^*(t)$. Due to space constraints, we note without proof that any equilibrium arrival profile $\textbf{F}$ is absolutely continuous. We now derive the equilibrium arrival profile illustrated in Figure \ref{fig:singledist}. Denote $\g = \a/(\a+\b)$.
\begin{theorem} \label{thm:single-equilibrium-dist}
Let $0=T_{s,1}, \leq T_{s,2} \leq \cdots \leq T_{s,K} $. Assume that $T_{s,K}< \frac{1 + \sum_{k=1}^K \mu_k T_{s,k}}{\sum_{k=1}^K \mu_k}$. Then, the unique equilibrium arrival profile is $F^* = (F_1^*, \cdots, F_K^*)$, where
\(
F_l^* = \frac{p_{0,l} (t+T_{0,l})}{T+T_{0,l}}
\)
with support $[-T_{0,l},T]$, where
\(
T = \frac{1 + \sum_{k=1}^K \mu_k T_{s,k}}{\sum_{k=1}^K \mu_k}\) and \(-T_{0,l} = (1-\frac{1}{\g} ) T + \frac{T_{s,l}}{\g},
\)
and the equilibrium routing probabilities are given by
\(
p^*_{0,l} = \frac{\mu_l}{\sum_{k=1}^K \mu_k} ( 1 - \sum_{k \ne l} \mu_l (T_{s,l} - T_{s,k}) ).
\)
\end{theorem}
\Proof
Note that the cost function is unbounded as $t$ goes to $\pm\infty$. Thus, at equilibrium the arrival profile must have bounded support. Let the support of the arrival profile to queue $l$ be $[-T_{0,l},T]$. Now, at equilibrium, the cost of arriving at queue $l$ is the same at any time in this arrival interval. Thus,
\(
C_l(T) = C_l(-T_{0,l}),
\)
from which we get 
\begin{equation} \label{fluid-frac}
p_{0,l} = \g \mu_l (T + T_{0,l}).
\end{equation}

Next, the equilibrium expected cost of arrival is the same at any queue and at any time in their respective arrival intervals. From Lemma \ref{lem:equilibrium-end-time}, we know that the time of last arrival at any queue is the same for all queues. Thus,  
 \( C_l(T) = C_k(T),\) for any $l,k$, and using \( \sum_{l \ne k} p_{0,l} = 1 - p_{0,k} \), we get
 \( \sum_{l \ne k} \mu_l ( \frac{p_{0,k}}{\mu_k} + T_{s,k} - T_{s,l} ) = 1 - p_{0,k}, \) rearranging which, we get that the equilibrium probability of routing to queue $k$ upon arrival is 
\(
p_{0,k} = \frac{\mu_k}{\sum_{l=1}^K \mu_l} ( 1 - \sum_{l \ne k} \mu_k (T_{s,k} - T_{s,l}) ).
\)

Now, from Lemma \ref{lem:idle-equilibrium}, we have that
\(
\mu_l(T - T_{s,k}) = p_{0,k}
\)
since the population size has been normalized to 1. Substituting for $p_{0,k}$, we get
\(
T = \frac{1 + \sum_{k=1}^K \mu_k T_{s,k}}{\sum_{k=1}^K \mu_k}.
\)
Now, it follows from equation \eqref{fluid-frac} that
\(
-T_{0,l} = T - \frac{p_{0,l}}{\g\mu_l}.
\)
Substituting for $T$ and $p_{0,l}$ we get 
\(
-T_{0,l} = \frac{1}{\sum_{k=1}^K \mu_k} \left ( 
(1-\frac{1}{\g} ) (1
+ \sum_{k=1}^K \mu_k T_{s,k} ) + \frac{T_{s,l}}{\g} \sum_{k=1}^K \mu_k \right ) \) which simplifies to   
\( -T_{0,l} =  (1-\frac{1}{\g} ) T + \frac{T_{s,l}}{\g}. \)

Finally, equating the cost of arrival at queue $l$ at any time $t$ with that at $T_{0,l}$ gives
\(
(\alpha + \beta) \frac{F^*_l(t)}{\mu_l} - \alpha t = \alpha T_{0,l},
\)
which yields
\(
F^*_l(t) = \mu_l \g (t+T_{0,l}) = \frac{p_{0,l}(t + T_{0,l})}{T + T_{0,l}},
\)
an equilibrium arrival profile at queue $l$.

We now argue uniqueness. First, note that for a given $T_{s,l}$, the terminal service time $T$ is unique. Let $F$ be another equilibrium profile with support $\mathcal{T}_i$ for $F_i$, where we can take $\mathcal{T}_i = (-\infty,T]$. Now, the cost of arriving at $T$ is $C_{F_i}(T) = \beta T = C_{F_i}(t)$ for each $t \in \mathcal{T}_i$. This is the same as under the profile $F^*$. Thus, $F_i(t) = F_i^*(t)$ on $\mathcal{T}_i \cap [-T_{0,i},T]$. Now, since $F_i^*(-T_{0,i})=0$ and $F_i^*(T)=p_{0,i}$, $F$ has total measure $1$ at $T$ and is absolutely continuous, it follows that $F_i(t) = F_i^*(t)$ on $[-T_{0,i},T]$. 
\EndProof

\noindent \textbf{Remarks.} 1. Note that we assume $T_{s,K} < T$ for convenience. Suppose $T_{s,l} > \frac{1 + \sum_{k=1}^{l-1} \m_k T_{s,k}}{\sum_{k=1}^{l-1} \m_k}$ for some queue $l \in \{2,\cdots,K\}$, then at equilibrium no users would arrive at queues $\{l,\ldots,K\}$. 
%\noindent \textbf{Remark.} We note that at equilibrium the number of queues that serve users used to serve the arriving users minimizes the service completion time $T$. To see this, suppose that $T_{s,L+1} < \frac{1+\sum_{k=1}^L \mu_k   T_{s,k}}{\sum_{k=1}^L \mu_k}$ and that there are no arrivals to queue number $L+1$ at equilibrium. Then, an arrival at $T_{s,L+1}$ can arbitrarily reduce its cost by switching to queue $L+1$. Thus, queue $L+1$ must be used at equilibrium. Further, let $T_{L+1} = \frac{1+\sum_{k=1}^{L+1} \mu_k   T_{s,k}}{\sum_{k=1}^{L+1} \mu_k}$ and $T_L = \frac{1+\sum_{k=1}^L \mu_k   T_{s,k}}{\sum_{k=1}^L \mu_k}$. Then, $T_{L+1} < T_{L}$ if and only if $T_{s,L+1} < T_L$ and the number of queues used at equilibrium minimizes the service completion time. Thus, fluid chooses to arrive at a queue in equilibrium so long as the service completion time is reduced by joining that queue.

%------------------------------------------------------
\subsection{Price of Anarchy} \label{sec:single-poa}
Define the {\em social cost} of arrival profile $F$ as
\(
J(F) = \sum_{k=1}^K \int C_{F_k}(t) dF_k(t).
\)
Let $J_{opt}$ denote the optimal social cost over all arrival profiles, and $J_{eq}(F^*)$ the social cost at equilibrium $F^*$. It is to be expected that  $J_{eq}(F^*)$ will be greater than $J_{opt}$. The inefficiency of the equilibrium arrival profile can be characterized by the {\em price of anarchy (PoA)},
\(
  \eta = \sup_{F^*} \frac{J_{eq}(F^*)}{J_{opt}},
\)
where the supremum is over all equilibria. Note that here the equilibrium arrival profile is unique.
\begin{theorem}
The price of anarchy of the network concert queueing game is given by
%\begin{equation} \label{pe:213}
\[
\eta = \frac{2(1 +  \sum_{k=1}^K \mu_k T_{s,k})}{\bigg ( 1 +    \sum_{k=1}^K  \sum_{l=1}^K \mu_k \mu_l T_{s,l}(T_{s,k} - T_{s,l})
    + 2 \sum_{k=1}^K \mu_k T_{s,k}  \bigg )}.
\]
%\end{equation}
\end{theorem}
\Proof
Let the equilibrium cost at queue $l$ be $C_l(t) = c \equiv \a T_{0,1}$ for all $t \in [-T_{0,l},T]$. The equilibrium social cost under profile $F^*$ is given by
\(
J_{eq} = \sum_{k=1}^K \int C_{F_k}(t) dF_k(t) = c \sum_{k=1}^K \int dF_k(t) = c (p_{0,1} + \cdots + p_{0,K}) =  c.
\)
Substituting for $c$, we have
\(
J_{eq} = \alpha T_{0,1} = \beta \frac{1 + \sum_{k=1}^K \mu_k T_{s,k}}{\sum_{k=1}^K \mu_k}.
\)

Now, the socially optimal outcome would be for each non-atomic user to arrive just at the instant of service, with zero waiting. In this case, the instantaneous cost would be 
\( C_{opt}(t) = \beta t \).
Thus, the optimal arrival profile is given by
\[
dF_{opt}(t) = \sum_{k=1}^l \mu_k dt, \quad ~\text{for}~T_{s,l} < t \leq T_{s,l+1}.
\]
 It is straightforward to see that the time of last arrival (and service) is $T_{opt} = T =  \frac{1 + \sum_{k=1}^K \mu_k T_{s,k}}{\sum_{k=1}^K \mu_k}$.
%\begin{cases} & \mu_1 dt  \quad ~\text{for}~0 \leq t \leq T_{s,2}, \\ & (\mu_1 + \mu_2) dt \quad ~\text{for}~T_{s,2} < t \leq T_{s,3},\\ & \vdots\\ & (\sum_{k=1}^K \mu_k) dt \quad ~\text{for}~T_{s,K} < t \leq T.  \end{cases}
From this, the optimal social cost can be computed as
\begin{eqnarray*}
J_{opt} &=&  \int_{0}^{T_{s,2}} \beta \mu_1 t dt + \int
_{T_{s,2}}^{T_{s,3}} \beta (\mu_1 + \mu_2) t dt + \cdots +
\int_{T_{s,K}}^T \beta \bigg (\sum_{k=1}^K \mu_{k} \bigg ) t dt,\\
&=& \frac{\beta}{2}  \bigg ( T^2 \sum_{k=1}^K \mu_k - \sum_{k=1}^K
\mu_k T_{s,k}^2 \bigg )
% &=& \frac{\beta}{2 \sum_{k=1}^K \mu_k} \bigg ( 1 + \bigg (
% \sum_{k=1}^K \mu_k T_{s,k} \bigg )^2 + 2 \sum_{k=1}^K \mu_k
% T_{s,k} - \sum_{k=1}^K \mu_k \sum_{k=1}^K \mu_k T_{s,k}^2 \bigg ),\\
= \frac{\beta}{2 \sum_{k=1}^K \mu_k} \bigg ( 1 + \sum_{k=1}^K
\sum_{l=1}^K \mu_k \mu_l T_{s,l}(T_{s,k} - T_{s,l}) + 2
\sum_{k=1}^K \mu_k T_{s,k} \bigg ).
\end{eqnarray*}
Using this along with the expression for $J_{eq}$ derived above, we get the expression for $\eta$.
%Thus, the price of anarchy is  \[ \eta = \frac{2(1 +  \sum_{k=1}^K \mu_k T_{s,k})}{ \begin{split} & \bigg ( 1 +     \sum_{k=1}^K  \sum_{l=1}^K \mu_k \mu_l T_{s,l}(T_{s,k} - T_{s,l})\     & + 2 \sum_{k=1}^K \mu_k T_{s,k}  \bigg )\end{split}}. \]
\EndProof

%The following corollary shows that the price of anarchy is upper bounded by 2.
\begin{corollary}\label{coro:poa-singlepopulation}
 The price of anarchy $\eta$ is upper bounded by 2.
 \end{corollary}
\Proof
  We will show that $J_{eq} \leq 2 J_{opt}$. Consider the difference $J_{eq} - 2 J_{opt} $ and simplify the expression to obtain
\(
J_{eq} - 2 J_{opt} =  \frac{\beta}{\sum_{k=1}^K \mu_k} \left (\sum_{k=1}^K \mu_k \sum_{k=1}^K  \mu_k T_{s,k}^2 - \sum_{k=1}^K\mu_{k} T_{s,k} \left ( \sum_{k=1}^K \mu_k T_{s,k} + 1 \right ) \right ).
\)
Recalling that $T = \frac{1+\sum_{k=1}^K \mu_{k} T_{s,k}}{\sum_{k=1}^K \mu_{k}}$, we can replace the last term on the R.H.S. above and simplify to get
\(
J_{eq} - 2 J_{opt} = \frac{\beta}{\sum_{k=1}^K \mu_k}\sum_{k=1}^K \mu_k \left ( \sum_{k=1}^K \mu_{k}T_{s,k} (T_{s,k} - T) \right ). 
\)
Now, we know from the statement of Theorem \ref{thm:single-equilibrium-dist} that $T > T_{s,k} \,\, \forall k$. Therefore, it follows that $J_{eq} \leq 2 J_{opt}$. 
 \EndProof
\noindent \textbf{Remarks.} 2. It is easy to see that the upper bound is achieved if $T_{s,k}^* = 0$. This is not surprising, as a set of parallel queues that start service at the same instant operate like a single server queue with effective service capacity $\sum_{k=1}^K \m_k$. \\
%Now, suppose that the queues start service at the same instant $\tau < \infty$. Then, simplifying the expression for the PoA, we obtain \( \eta = 1 + \frac{1}{1+\t \sum_{k=1}^K\m_k}. \) Let $K \rightarrow \infty$ then, clearly, $\eta \rightarrow 1$. If there are an infinite number of queues, then any arriving user will always find service capacity available in the network. Not surprisingly, the optimal and equilibrium arrival behaviors have the same social cost in this case. 
3. Surprisingly though, Corollary \ref{coro:poa-singlepopulation} implies that staggering the start times of the queues can reduce the PoA (even though it may increase the social cost), and induce arrival behavior closer to  the optimum. \\
4. As a special case, consider that all queues have the same service rate $\frac{\m}{K}$ and start at times spaced $\t$ apart. Then, the PoA expression reduces to
\begin{equation} \label{eta-special-case}
\eta = \frac{2 + \m \t (K-1)}{1 + \m \t (K-1) - \frac{\m^2 \t^2}{12}(K^2-1)}.
\end{equation}
An easy lower bound on this expression follows from the fact that $1 + \m \t (K-1) > 1 + \m \t (K-1) - \frac{\m^2 \t^2}{12}(K^2-1)$, which after simple algebra (and using some elementary facts)  yields
\(
2 > \eta > 1 + \frac{1}{1 + \m \t (K-1)} > 4/3.
\)
\section{The Network Concert Game with Multiple Populations}\label{sec:multiple} 
%\begin{figure}[h] \label{fig:multiple} \centering\includegraphics[scale=0.75]{../Figures/fig-multiple.pdf} \caption{Multiple populations arriving at a parallel queue network. Each population has its own delay risk characteristic $(\alpha_i,\beta_i)$.} \end{figure}

We now consider multiple populations of users arriving at a $K$ parallel queueing network. Let $\sN = \{1, \cdots, N\}$ be the set of arriving populations, each with risk characteristic $(\alpha_i,\beta_i)$. Denote $\g_i = \alpha_i/(\alpha_i + \beta_i)$. Let the service start times be $0 = T_{s,1} \leq \cdots \leq T_{s,K}$ with mean service rates $\m_k$. Recall, from Section \ref{sec:strategic-arrivals}, that the (fluid limit of the) expected cost of arrival for a population $i$ user to queue $k$ at time $t$ is given by $C_{i,k}(t) = (\a_i + \b_i)W_{i,k}(t) + \b_i t$, which as earlier, is constant over the arrival interval, and same across all the queues in the network. 

\begin{figure}[h]\centering {\includegraphics[scale=0.6]{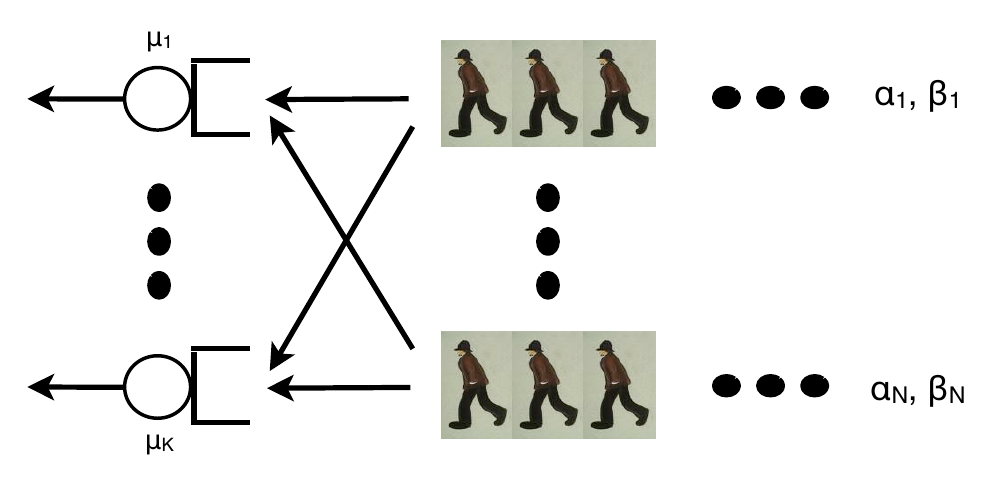}}\caption{Arrivals from multiple populations into multiple queues.}\label{fig:multipledist}\end{figure}

The following Lemma shows an interesting \textit{self-organization property} at equilibrium.
\begin{lemma} \label{lem:order-arrival}
% Consider a single server queue offering service rate $\mu$. Suppose there are $N$ populations with $\g_1 < \cdots < \g_N$. Then, at equilibrium the populations have to arrive in order of $\g_i$ and over disjoint intervals. 
Suppose that $\g_1 < \g_2 \cdots < \g_N$. Then, at equilibrium population $i$ users arrive before population $j$ users for $i < j$. Furthermore, the arrivals are over disjoint intervals, without any gaps.
\end{lemma}
%(The proof for the parallel network case follows along the lines of Lemma 2 of \cite{JaJuSh2010a}. )
\Proof
Let $T_{s,k} = 0$, for all queues $k$. The general case will follow easily from the ensuing argument. First note that there can be no gaps in any equilibrium arrival profile, $F_k = \sum_{i=1}^N F_{i,k}$. If there were, then, any arriving non-atomic user right after the gap can arbitrarily improve its cost by arriving just before such a gap, implying this arrival profile is not in equilibrium. Now, the cost of arriving at queue $k$ is constant, for a given population $i$ over the arrival interval. Differentiating $C_{i,k}(t)$, we have 
\(
0 = (\a_i + \b_i)(\frac{1}{\mu_{k}} \frac{\partial F_{i,k}(t)}{\partial t}) - \a_i.
\)
Solving the equation for $f_{i,k}$, the arrival density, we have $f_{i,k}(t) = \mu_k \gamma_i$.

Now, let $(t_1,t_3)$ be an arbitrary interval, and suppose that $t_1 < t_2 < t_3$, such that some population $j$ users arrive in $(t_1,t_2]$ and only population $i$ users arrive in $(t_2,t_3)$. Consider the cost of arrival for population $j$, $C_{j,k}(t) = (\a_j + \b_j)\frac{F_k(t)}{\m_k} - \a_j t$, for $t \in (t_1,t_3)$. As $(t_1,t_2]$ is in the support of $F_{j,k}$, it follows that the cost of arrival is constant over this interval, and it can be evaluated at $t_2$. Now, let $\e > 0$ be small enough so that $t, t+\e \in (t_2,t_3)$. Evaluating the cost of arrival at these points and taking the difference of the resulting expressions we obtain
\(
C_{j,k}(t+\e) - C_{j,k}(t)  = (\a_j + \b_j) \left( \frac{\sum_{l=1}^N F_{l,k}(t+\e) - \sum_{l=1}^N F_{l,k}(t)}{\m_k} \right ) - \a_j \e.
\)
By assumption, there are only arrivals from population $i$ in the sub-interval $(t_2,t_3)$, and it follows that $F_{l,k}(t+\e) = F_{l,k}(t)$ for all $l \neq i$. This implies that
\(
C_{j,k}(t+\e) - C_{j,k}(t) = (\a_j + \b_j) \frac{(F_{i,k}(t+\e) - F_{i,k}(t))}{\m_k} - \a_j \e.
\)
Divide through by $\e$ and let $\e \rightarrow 0$ to obtain
\(
C^{'}_{j,k}(t) = (\a_j + \b_j) \frac{f_{i,k}(t)}{\m_k} - \a_j,
\)
where $f_{i,k}(t)$ is the density of the arrival profile $F_{i,k}(t)$, which was shown to be $f_{i,k}(t) = \m_k \g_i$, for $t \in (t_2,t_3) \subset \text{Supp} F_{i,k} $. Substituting for $f_{i,k}$ it follows that
\(
C^{'}_{j,k}(t) = (\a_j + \b_j)(\g_i - \g_j).
\)
By assumption, $\g_i < \g_j$, so that $C^{'}_{j,k}(t) < 0$. This implies that the cost of arrival for population $j$ is strictly decreasing over the interval $(t_2,t_3)$, and it must be less than $C_{j,k}(t_2)$. Clearly, $(t_1,t_2]$ cannot be in the support of $F_{j,k}$ in equilibrium. This implies that population $j$ cannot arrive before population $i$, if $i < j$.
\EndProof
%Since arrivals to queues in parallel must end at the same instant by Lemma \ref{lem:equilibrium-end-time} it still follows that the arrivals occur over disjoint intervals. This means that we can continue to treat the arriving users as a single homogeneous class in the parlance of standard queueing theory.

\subsection{Arrival Distribution} \label{sec:multiplearrival}
\begin{figure}[t]
\centering
\includegraphics[scale=1]{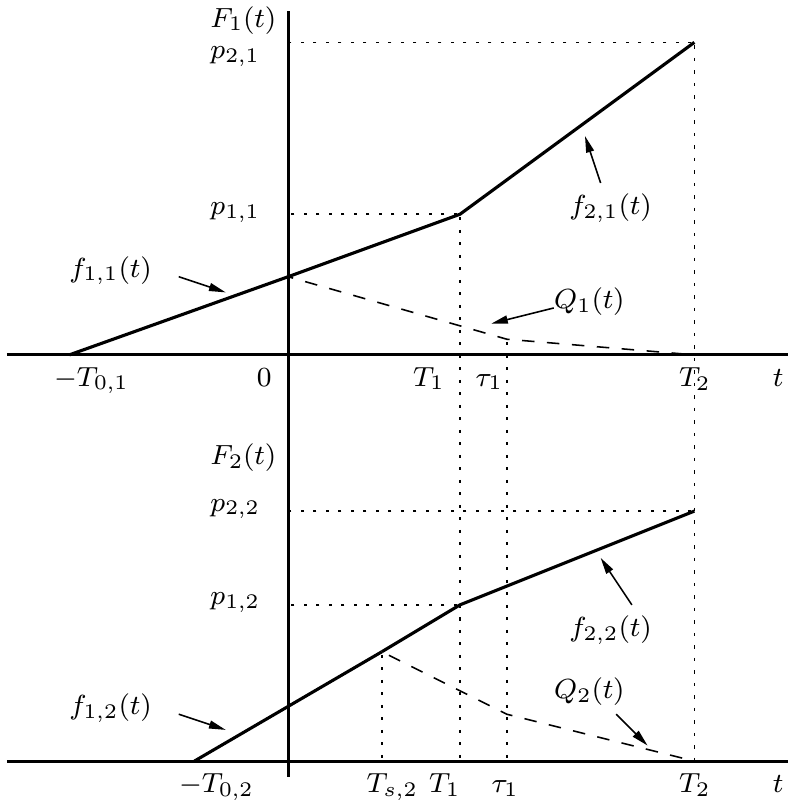}
\caption{Two parallel queues, and two arriving populations with $\g_1 < \g_2$. Population $1$ arrives over $[-T_{0,1},T_1]$ at queue 1, and over $[-T_{0,2},T_1]$ at queue 2. However, population $1$ need not be served completely until time $\t_1$ at either queue. Population 2 arrives over $[T_1,T_2]$ and is completely served at time $T_2$.  }
\label{fig:contrast}
\end{figure}
% Before proving the existence and uniqueness of the equilibrium arrival distribution, we establish some notation.
Denote by $-T_{0,k}$, the time of first arrival into queue $k$, and $T_0 := -T_{0,1}$, the time of the very first arrival to the network. Then,  from Lemma \ref{lem:order-arrival}, users from population $i$ arrive in the interval $[T_{i-1}, T_i]$. Let $\t_i$ denote the time the last user of population $i$ is served, and $\t_0=0$. Obviously, $T_i \leq \t_i$, with equality only if $i = N$ since at equilibrium the last (non-atomic) user to arrive into the network has no incentive to arrive before his service time. Figure \ref{fig:contrast} provides a simple illustration of this phenomenon, for two arriving populations at two parallel nodes. 

%Before proving the existence and uniqueness of the equilibrium arrival distribution, we establish some notation. Users from population $1 < l \leq N$ arrive over the disjoint intervals $[T_{l-1}, T_l]$, where $T_l$ is the time of the very last arrival from population $l$. If $l =1$ then the arrivals occur over the interval $[-T_{0,1}, T_1]$, where, as before, $T_1$ is the time of the very last arrival from population $1$, and $-T_{0,1}$ is the time of the very first arrival from population 1 to the very first queue in operation. Let $J_{l}$ be the set of queues that commence service in the interval $[T_{l-1},T_l]$, i.e., if $1 < l \leq N$, $J_l = \{1 \leq k \leq K : T_{s,k} \in [T_{l-1}, T_{l}] \}$ and if $l =1$ $J_1 = \{1 \leq k \leq K : T_{s,k} \in [-T_{0,1},T_1]\}$. Then the servers serving population $l$ will be $\bigcup_{i=1}^l J_i$. Also, queues in $J_l$ will serve all populations $l \leq n \leq N$.

Define $J_i = \{1 \leq k \leq K : T_{s,k} \in [\tau_{i-1}, \tau_{i}] \}$. Then, population $i$ users are served by the queues $\sJ_i = \bigcup_{j=1}^i J_j$, where queues $J_i$ first serve population $i$ before any other.  Consider $l \in J_i$ with aggregate arrival profile $F_{l}(t) = \sum_{n = i}^N F_{n,l} (t)$, where by Lemma \ref{lem:order-arrival}, $F_{n,l}$ has support $[T_{n-1},T_n]$. Note that $\sum_{l \in \sJ_i} p_{i,l} = \sum_{j=1}^i \sum_{k \in J_j} p_{i,k} = 1$. We can now derive the equilibrium arrival profile for each population. % (illustrated in Figure \ref{fig:contrast} for the case of two arriving populations into)

\begin{theorem} \label{thm:multiple-equilibrium-dist}
Suppose $\g_i < \g_{i+1}$, $\forall i$ and $T_{s,k} < T_{s,k+1}$, $\forall k$. Then, the unique equilibrium arrival profile for population $i$ at queue $k \in J_j$, $j < i$, is $dF^*_{i,k}(t) = \g_i \mu_{k}, \quad \forall t \in [T_{i-1},T_i]$, and at queue $l \in J_i$ is $dF^*_{i,l}(t) = \g_i \mu_{l} \quad \forall t \in [-T_{0,l},T_i]$ where $T_N = \frac{1}{\sum_{i=1}^N \sum_{k \in J_i} \mu_{k}} \left (N + \sum_{i=1}^N \sum_{k \in J_i} \mu_{k} T_{s,k} \right )$, and $T_{i-1} = T_{i} - \frac{p_{i,k}}{\g_i \mu_{k}}$ for $i=1,\cdots,N-1$, $k \in J_j$ and $j < i$. For $l \in J_i$, the arrival interval is $[-T_{0,l},T_i]$, where $-T_{0,l} = T_{i} - \frac{p_{i,l}}{\g_i \mu_{l}}$. 

Furthermore, equilibrium routing probability for $l \in J_i, i \geq 1$  is 
\begin{eqnarray} 
\label{routing-prob-l}
p_{i,l}= \frac{\mu_{l}}{\sum_{j=1}^{i} \sum_{k \in J_j}
  \mu_{k}} \bigg ( i - \sum_{j=1}^{i} \sum_{k \in J_j} \mu_{k}
(T_{s,l}-T_{s,k}) \bigg ),
\end{eqnarray}
and for $k \in J_j$, $j < i$, and $i \geq 2$ is
\begin{eqnarray} 
\label{routing-prob-k}
p_{i,k} = \frac{\mu_{k}}{\sum_{j=1}^i \sum_{k \in J_j}\mu_k}
\bigg ( 1 - \sum_{l \in J_i} \frac{\mu_{l}}{\mu_{k}}
(\sum_{q=j}^{i-1} p_{q,k}) - \sum_{l \in J_i} \mu_{l}
(T_{s,k} - T_{s,l}) \bigg ).
\end{eqnarray}
\end{theorem}
\Proof
Note that population $i$ is served by queues $J_l, l \leq i$. The expected cost for a population $i$ user to arrive at queue $k \in \sJ_i$ is $C_{i,k}(t) = (\alpha_i + \beta_i) W_{k}(t) + \beta_i t$ for $t \in [T_{i-1},T_i]$ for $k \in J_l, l<i$ ,  and for $t \in [-T_{0,k},T_{i}]$ for $k \in J_i$. Recall that, $F_{k}(t) = \sum_{j=i}^{N} F_{j,k}(t)$, where $F_{i,k}(t)$ has support $[-T_{0,k}, T_i]$ and $F_{jk}(t)$ has support $[T_{j-1},T_j]$ for each $j>i$. The virtual waiting time process at queue $k$ is given by $ W_{k}(t) = \frac{1}{\mu_{k}} F_{k}(t) - (t - T_{s,k})$. 

Now, note that at equilibrium, the expected cost for a population $i$ user to arrive at any queues $k,l \in J_i$ has to be the same. Thus, from $C_{i,k}(T_i) = C_{i,l}(T_i)$, we get that
\begin{equation} \label{lll}
\frac{p_{i,k}}{\mu_{k}} + T_{s,k} = \frac{p_{i,l}}{\mu_{l}} + T_{s,l}.
\end{equation}
Let $p \leq q < i$ be two populations that arrive prior to population $i$. Then, for any $k \in J_p$ and $m \in J_q$, $k<m$ and at equilibrium the expected cost of arriving into queues $k$ and $m$ for a population $i$ user has to be the same. Thus, from $C_{i,k}(T_i) = C_{i,m}(T_i)$, we get 
\begin{equation}\label{lkm}
\frac{1}{\mu_{k}}(p_{p,k} + p_{p+1,k} + \cdots +
p_{i,k}) + T_{s,k} = \frac{1}{\mu_{m}}(p_{q,m} + p_{q+1,m} + \cdots +
p_{i,m}) + T_{s,m}.
\end{equation}

First, consider $p < q$. Then, for a population $q$ user, $C_{q,k}(T_q) = C_{q,m}(T_q)$ for queues $k,m$ as above, which yields $\frac{1}{\mu_{k}}(p_{p,k} + p_{p+1,k} + \cdots + p_{i,k}) + T_{s,k} = \frac{1}{\mu_{m}}(p_{q,m}) + T_{s,m}$. Substituting this into \eqref{lkm} we have, by induction that
\begin{equation} \label{lkm-final}
\frac{p_{i,m}}{\mu_{m}} = \frac{p_{i,k}}{\mu_{k}}, \quad\forall k \in J_p, ~ m \in J_q, ~p<q< i.
\end{equation}
Now consider $p = q$. Then, for any $k, l\in J_p$ we have $C_{i,k}(T_i) = C_{i,l}(T_i)$, implying $\frac{1}{\mu_{k}}(p_{p,k} + \cdots + p_{i,k}) + T_{s,k} = \frac{1}{\mu_{l}}(p_{p,l} + \cdots + p_{i,l}) + T_{s,l}$. It follows from \eqref{lll} that $\frac{p_{p,k}}{\mu_{k}} + T_{s,k}  = \frac{p_{p,l}}{\mu_{l}} + T_{s,l}$. Thus, by induction, we have
\begin{equation} \label{lkk}
\frac{p_{i,l}}{\mu_{l}} = \frac{p_{i,k}}{\mu_{k}}, \quad \forall k,l \in J_p, ~p<i.
\end{equation}
Next, let population $j < i$, and $k \in J_j$ and $l \in J_i$ are two queues that serve population $i$. Once again by the equilibrium conditions we have $C_{i,k}(T_i) = C_{i,l}(T_i)$. Simplifying the expression we obtain
\begin{equation} \label{lm}
p_{i,k} = \frac{\m_k}{\m_l} p_{i,l} + \m_k (T_{s,l} - T_{s,k}) - \sum_{p=j}^{i-1} p_{p,k}.
\end{equation}

Now,  we have \( \sum_{p=1}^i \sum_{m \in J_p} p_{i,m} = 1. \) Consider a $j < i$ and a $k \in J_j$ for $i\geq 2$. It follows that $\sum_{p=1, p \ne j}^i \sum_{m \in   J_p} p_{i,m} + \sum_{l \in J_j, l \neq k} p_{i,l} = 1 - p_{i,k}$. Substituting for $p_{i,m}$ and $p_{i,l}$ in terms of $p_{i,k}$ from \eqref{lkm-final}, \eqref{lkk} and \eqref{lm}, we get $p_{i,k}$ in (\ref{routing-prob-k}). Now, for an $l \in J_i$ for any $i \geq 1$, using \eqref{lll} and \eqref{lm} and substituting for $p_{i,k}$ and $p_{i,m}$ (respectively) in terms of $p_{i,l}$ in $\sum_{j=1}^{i-1} \sum_{m \in J_j} p_{i,m} + \sum_{k \in J_i, k \neq l } p_{i,k} = 1 - p_{i,l}$, we get $p_{i,l}$ in  (\ref{routing-prob-l}).

% Now,  we have \( \sum_{p=1}^i \sum_{m \in J_p} p_{i,m} = 1. \) Consider a $j < i$ and a $k \in J_j$ for $i\geq 2$. It follows that $\sum_{p=1, p \ne j}^i \sum_{m \in   J_p} p_{i,m} + \sum_{l \in J_j, l \neq k} p_{i,l} = 1 - p_{i,k}$. Substituting for $p_{i,m}$ and $p_{i,l}$ in terms of $p_{i,k}$ from \eqref{lkm-final} and \eqref{lkk}, we get $p_{i,k}$ in (\ref{routing-prob-k}). Now, for an $l \in J_i$ for any $i \geq 1$, using \eqref{lkm-final} and \eqref{lll} in $\sum_{j=1}^{i-1} \sum_{m \in J_j} p_{i,m} + \sum_{k \in J_i, k \neq l } p_{i,k} = 1 - p_{i,l}$, we get $p_{i,l}$ in  (\ref{routing-prob-l}).

% Finally, consider the last population $N$. Note that there are two possibilities. First, suppose there is a set $J_N$ of queues that commence service in $[\t_{N-1},T_N]$ and serve only population $N$. In this case the, same equations as above hold. Second, suppose that no new queues start service in this interval. Then, it is easy to derive the fact that $\forall l \in \{ 1, \cdots, N-1\}$ and $\forall i_l \in J_l$, $p_{N,i_l} = \frac{\mu_{i_l}}{\sum_{k=1}^{N-1} \sum_{i_k \in J_k} \mu_{i_k}}$.

We now derive the equilibrium arrival distributions for each population to each serving queue. First, recall that the cost function for population $i$ at queue $k \in J_j, j < i$, is given by $C_{i,k}(t) = (\alpha_i + \beta_i) \left ( \frac{1}{\mu_{k}} F_{k}(t) + T_{s,k} \right )- \alpha_i t$ $\forall t \in [T_{i-1},T_i].$ Differentiating this and recalling that at equilibrium the cost is constant over the arrival interval, and that $F_{k}(t) = \sum_{p=j}^{i-1} p_{p,k} + F_{i,k}(t)$, we have $dF_{i,k}(t) = \g_i \mu_{k} \quad \forall t \in [T_{i-1},T_i]$. Now, for queue $k \in J_i$, the users arrive over the interval $[-T_{0,k},T_i]$. Again, differentiating the cost function $C_{i,k}(t)$, we have $ dF_{i,k}(t) = \g_i \mu_{k} \quad \forall t \in [-T_{0,k},T_i]$.

Finally, we can derive the support of these distributions by backward recursion. Note that for population $N$, $\forall k \in J_N$ $\mu_{k} (T_N - T_{s,k}) = p_{N,k}$. Substituting for $p_{N,k}$ from \eqref{lll}, we get $T_N = \left ( N + \sum_{j=1}^N \sum_{k \in J_j} \mu_{k} T_{s,k} \right )/\left(\sum_{j=1}^N \sum_{k \in J_j} \mu_{k}\right)$.%  On the other hand, suppose that $J_N = \emptyset$. The time by which population $l$ is served is given by $\tau_l = \frac{l + \sum_{k=1}^l \sum_{i_k \in     J_k} \mu_{i_k} T_{s,i_k}}{\sum_{k=1}^l \sum_{i_k \in J_k}   \mu_{i_k}}$. The proof of this fact is easily verified by recursion. Note that this is not the same as $T_l$, the time at which the last population $l$ arrival occurs. Now, it follows that the very last population $N$ will be completely served by $T_N = \frac{N + \sum_{k=1}^{N-1} \sum_{i_k \in     J_k} \mu_{i_k} T_{s,i_k}}{\sum_{k=1}^{N-1} \sum_{i_k \in J_k} \mu_{i_k}}$.

Next, at equilibrium, we must have $C_{i,k}(T_i) = C_{i,k}(T_{i-1})$ $\forall k \in J_j, j < i$, from which we get
\( T_{i-1} = T_{i} - \frac{p_{i,k}}{\g_i \mu_{k}}. \)
Note that we need to use $j < i$ in order to obtain the recursive definition of $T_{i-1}$, since $C_{i,k}(t) < C_{i,l}(t)$ on $[T_{i-1},-T_{0,l}]$, for $l \in J_i$; that is, there are no arrivals from population $i$ at queue $l \in J_i$ on this sub-interval. Finally, population $i$ users arrive at queue $k \in J_i$ in the interval $[-T_{0,l},T_i]$. Thus, at equilibrium we must have $C_{i,k}(-T_{0,k}) = C_{i,k}(T_i)$, from which we obtain $-T_{0,k} = T_i - \frac{p_{i,k}}{\g_i \mu_{k}}$. 

The proof of uniqueness now follows that of Theorem \ref{thm:single-equilibrium-dist} and we omit it for brevity.
\EndProof

\noindent \textbf{Remarks.} 1. Theorem \ref{thm:multiple-equilibrium-dist} shows that at equilibrium the arriving populations self-organize in ascending order of $\g_i$ % and that the populations arrive completely before they have been fully served out
% . The second population arrives immediately after the first (and so on), 
and there are no gaps in the arrival profile. Further, the queues operate at full capacity till all arriving users have been served.

%-------------------------------------------------------------
\subsection{Price of Anarchy} \label{sec:multiple-poa}

We now compute the \textit{price of anarchy} for the multiple populations case. % We will first show how this expression can be derived in terms of the known parameters of the network; viz., the service rates offered and the service start times of the queues. The resulting expressions are complex and offer no insight. So rather than belabor the algebra, we illustrate the price of anarchy for a special case.
% We start with the equilibrium social cost. Recall that the set of queues that serve population $i$ is $\sJ_i = \cup_{j=1}^i J_j$. As noted in Section \ref{sec:single-poa}, the equilibrium social cost is the total cost of a unit fluid from population $i$ arriving into the queues serving that population, i.e., for 
Define the social cost at equilibrium with arrival profile $F$, as $J(F) = \sum_{i=1}^N J_{i}(F)$, where $J_{eq,i} = \sum_{j = 1}^{i-1} \sum_{k \in J_j} \int_{T_{i-1}}^{T_i} C_{i,k}(t) dF_{i,k}(t) + \sum_{l \in J_i}\int_{-T_{0,l}}^{T_i} C_{i,l}(t) dF_{i,l}(t)$.

At equilibrium, the  cost of arrival for population $i$ is uniform over the support of its arrival profile, and moreover is the same at all queues that the population chooses to arrive at. %Let $C_l(t)$ denote $\min_{j \in \cup_{i=1}^lJ_i} C_{l,j}(t)$, the minimum cost in any operational queue faced by a population $i$ user if it were to arrive at time $t$.  
Thus, $C_{i,l}(t) = c_i$, some constant. Then, if $F$ is an equilibrium arrival profile
\(
J_{eq,i} = c_i \left( \sum_{j = 1}^{i-1} \sum_{k \in J_j} \int_{T_{i-1}}^{T_i} dF_{i,k}(t) + \sum_{l \in J_i}\int_{-T_{0,l}}^{T_i} dF_{i,l}(t) \right) = c_i \sum_{j = 1}^{l} \sum_{k \in J_j} p_{0,k} = c_i,
\)
since $\sum_{j = 1}^{i} \sum_{k \in J_j} p_{0,k}$ = 1. The aggregate equilibrium social cost is given by $J_{eq} = \sum_{i=1}^N J_{eq,i} = \sum_{i=1}^N c_i$.

Let $e_1(i)$ denote the ``first'' queue in $J_i$, i.e., the queue with the earliest service start time $T_{s,l}$, $l \in J_i$. At equilibrium we have $C_{i,e_1(i)} (T_i) \,=\, (\a_i + \b_i)  ( \frac{p_{i,e_1(i)}}{\mu_{e_1(i)}} + T_{s,e_1(i)}  ) - \a_i T_i \, \equiv \, c_i$, where, $p_{i,e_1(i)}$ is the fraction of population $i$ users routed to queue $e_1(i)$. Now, let $e_1(1)$ be the very first queue to start service (and serve population 1 first). Without loss of generality, let $T_{s,e_1(1)} = 0$. For population $i$ the cost of arrival at any queue in $\sJ_i$ is the same over the arrival interval, and it follows that $C_{i,e_1(i)}(T_i) = C_{i,e_1(1)}(T_i)$, which implies that $\frac{p_{i,e_1(i)}}{\mu_{e_1(i)}} + T_{s,e_1(i)} = \sum_{j=1}^i \frac{p_{j,e_1(1)}}{\mu_{e_1(1)}}$. Further, using the recursive definition of $T_i$, $T_i \,=\, T_N - \sum_{j=i+1}^N \frac{p_{j,e_1(1)}}{\gamma_j \mu_{e_1(1)}}$. Substituting for $\frac{p_{i,e_1(i)}}{\mu_{e_1(i)}} + T_{s,e_1(i)}$ and $T_i$ in $C_i(T_i)$, we obtain
\begin{equation} \label{social-cost-eq}
J_{eq} = \sum_{j=1}^N \alpha_j \bigg ( \frac{1}{\gamma_i} \bigg (\sum_{j=1}^i \frac{\mu_{e_1(i)}}{\mu_{e_1(1)}} p_{j,e_1(1)} \bigg ) - \mu_{e_1(i)} T_{s,e_1(i)}  -  \bigg ( T_N - \sum_{j=i+1}^N \frac{p_{j,e_1(1)}}{\gamma_j \mu_{e_1(1)}} \bigg ) \bigg ).
\end{equation}
Now, from Theorem \ref{thm:multiple-equilibrium-dist} we have an expression for $T_N$ in terms of the exogeneous parameters of the network. Substituting that into \eqref{social-cost-eq} we obtain an expression for $J_{eq}$ that is, unfortunately, quite messy for the general case. Below, we  illustrate this expression for a much simpler special case.

Next, we note that the optimal arrival profile would be for each non-atomic user to arrive right at the instant of service. In this case, there is no waiting time and the cost of arrival at time $t$ is $\beta_i t$, for a user of population $i$. Let $\pi : \sN \rightarrow \sN$ be a permutation on the set of populations such that $\beta_{\pi(1)} > \cdots > \beta_{\pi(N)}$. In the optimal arrival profile populations should arrive in the order $\pi(1), \pi(2), \cdots, \pi(N)$. 

A key observation is that since the sizeof each population is the same, the set of queues $\sJ_{i}$ that serve population $i$ at equilibrium, will now serve population $\pi(i)$. Let $\sJ^*_{i}$ be the set of queues that serve population $\pi(i)$ in the optimal arrival profile.
%For ease of notation, we will assume that $\beta_1 > \beta_2 > \cdots > \beta_N$ in the ensuing discussion. though it should be kept in mind that population $l$ here is really population $\pi(l)$. 
%Note that this simplification of notation does not affect the results we derive next. It will be shown that the social cost of optimal arrival behavior, for any population $l$, depends on the interval in which it arrives (modulo the term $\beta_l$). We show, in Lemma \ref{lem:serve-set}, that the arrival interval is in turn dependent on the size of the opulation/total volume of fluid that has arrived by the end of the arrival interval. ]]
%\textcolor{red}{(Since each population has the same size, the exact order of arrival does not matter in the computation of optimal social cost[??])}
This is because there are no gaps in the optimal and equilibrium arrival profiles. Thus, for given queue start time times, $T_{s,k}$, if population $i$ in the equilibrium arrival profile is replaced by population $\pi(i)$, the set of queues that served population $i$ will now serve population $\pi(i)$, whose users now arrive over $[T_{i-1},T_i]$, which is the equilibrium arrival interval for population $i$.

Thus, the optimal social cost is
\begin{equation} \label{social-cost-opt}
J_{opt} = \sum_{i=1}^N J_{opt, \pi(i)} = \frac{1}{2} \sum_{i=1}^N \beta_{\pi(i)} \bigg ( \sum_{j=1}^{i-1} 
\sum_{k \in J_j} \mu_{k} ((T_i^*)^2  - (T_{i-1}^*)^2)+\: \sum_{l
\in J_i} \mu_{l} ((T_i^*)^2 - T_{s,l}^2) \bigg ),
\end{equation}
%To derive $T_l^*$, first note that the set of queues that serve population $l > 1$ in the optimal behavior case is the same as the set of queues that serve population $l$ at equilibrium. To see this, recall that when the arrival behavior is optimal, population $l$ arrives at time $T^*_{l-1}$ and that there are some queues that have started service before this time. Obviously, this population will be served by queues other than the ones in service already, provided the new queues commence service before the unit fluid of population $l$ is served out. \textcolor{red}{However, recall from the remark after Theorem \ref{thm:single-equilibrium-dist} that this is precisely the condition that arrivals in equilibrium use to choose the set of queues to arrive at.} 
% Now, population $\pi(1)$, in optimal arrival behavior, starts arriving at time $0$ and up to 
% $T_1^* = \frac{1 + \sum_{i_1 \in J_1} \mu_{i_1}}{\sum_{i_1 \in J_1} \mu_{i_1}}$, which is the time in which one unit of fluid can be served. Population $\pi(2)$ arrives from $T_{1}^*$ to $T_2^*$ given by $T_2^* = \frac{2 + \sum_{k=1}^2 \sum_{i_k \in J_k} \mu_{i_k}
%   T_{s,i_k}}{\sum_{k=1}^2 \sum_{i_k \in J_k} \mu_{i_k}}$.
where $T_i = \frac{i + \sum_{j=1}^i \sum_{k \in J_j} \mu_{k}   T_{s,k}}{\sum_{j=1}^i \sum_{k \in J_j} \mu_{k}}$, and this is precisely the time at which population $i$ finishes service at equilibrium. We can substitute for $T_i$ in \eqref{social-cost-opt}, which yields a fairly complicated expression for $J_{opt}$ and the price of anarchy, $\eta = J_{eq}/J_{opt}$, in terms of the exogeneous parameters.

To get some insight into the price of anarchy, $\eta$, we illustrate it for a special case where the service rate offered by every queue is the same $\mu > 0$ and the start time of the $k$th queue is $\tau(k-1)$, for some $\tau > 0$. % Thus, the first queue starts at time $0$, the second queue at $\tau$ and so on.
Let the number of queues that serve the first $l$ populations to arrive be $K_l$. Then, the instant at which population $\pi(l)$ (or population $l$, at equilibrium) is served out is given by $T_l = \frac{l + \sum_{k=1}^{K_l} \mu \tau (k-1)}{\sum_{k=1}^l \mu}= \frac{l + \frac{\mu \tau}{2} K_l (K_l - 1)}{\mu K_l}$. Note that $K_l$ is unknown a priori, but can be easily calculated. % The number of queues that can serve the first population to arrive depends on the amount of fluid that the first server to commence operation, $e_1(1)$, can serve by $\tau$. That is, when $\mu \tau < 1$, there are at least two queues that serve population 1 (and hence all other populations).
Suppose $\mu \tau < 1$, and relax $K_l \in [1,K]$ to take real values. Then, the following Lemma shows that $T_l$ is a convex function of $K_l$. The optimal value of $K_l$ then is the nearest integer to the optimal real value computed. Let $[x]$ denote the nearest integer to the real number $x$.

\begin{figure}[ht] 
\centering
{\includegraphics[scale=0.3]{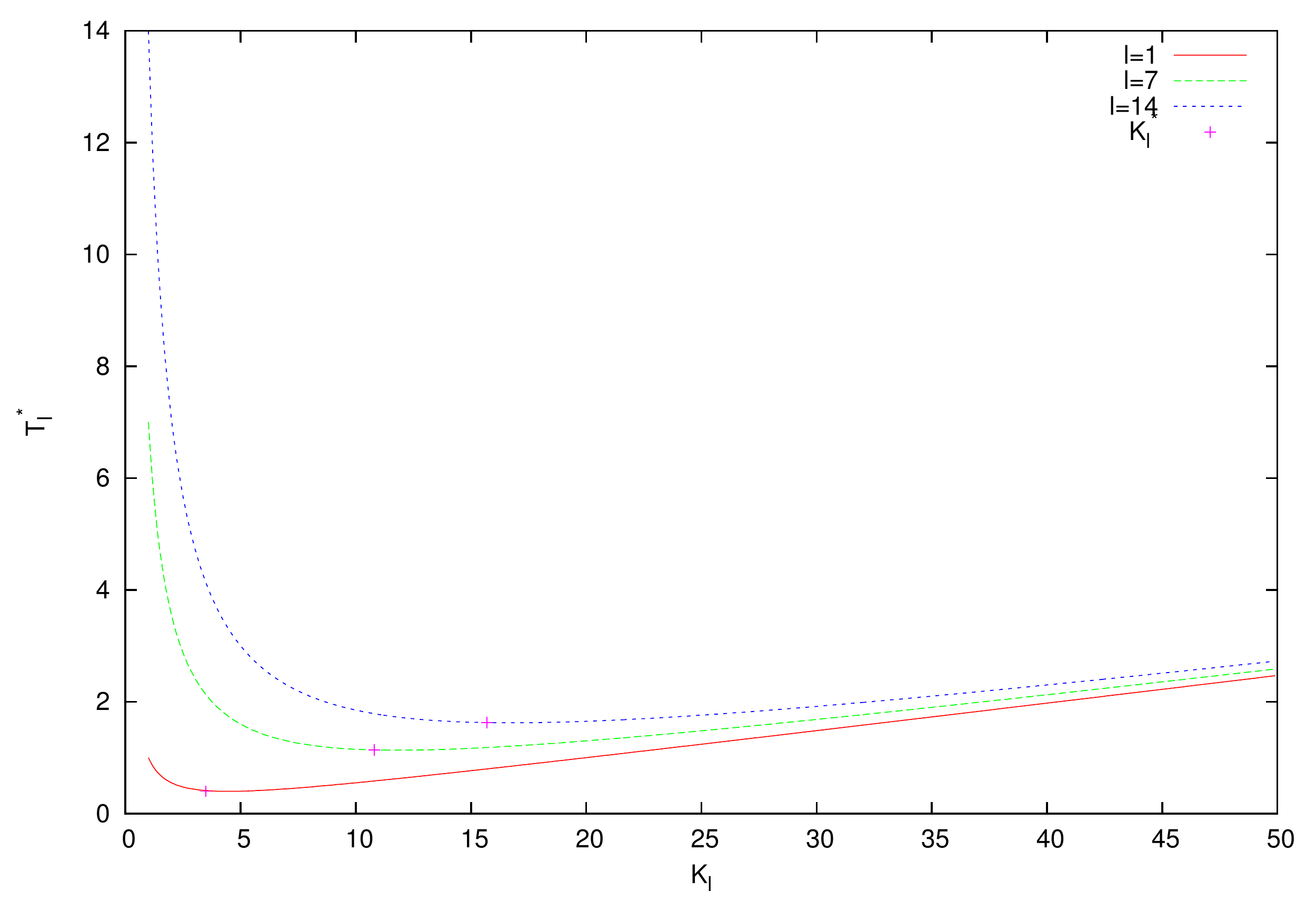}}
\caption{Terminal arrival time $T_l$ as a function of $K_l$, the number of queues that service population $l$, plotted for $l = 1,7,14$, and for $\mu \tau = 0.1$ and $\mu = 1.0$.}
\label{fig:vary_l}
\end{figure}
Figure 4 plots $T_l$ as a function of $K_l$, the number of queues that serve population $l$. Note that $T_l$ is a convex function of $K_l$ and has a minima. We establish this fact formally in Lemma \ref{lem:serve-set}.

\begin{lemma} \label{lem:serve-set}
Suppose $\mu \tau < 1$. % Consider $T_l$ to be a real-valued function of $K_l$. 
Then, $T_l$ is a convex  function of $K_l \in (1,K]$. Further, it achieves a minimum at $K^*_l = [\sqrt{\frac{2 l}{\mu \tau}}]$.
\end{lemma}
\Proof
From Theorem \ref{thm:multiple-equilibrium-dist} we have $T_l(k) = \frac{l}{\mu k} + \frac{\tau}{2} (k -1)$, for $k \in (1,K]$. Differentiating with respect to $k$ we obtain $\frac{\partial T_{l}(k)}{\partial k} = -\frac{l}{\mu k^2} + \frac{\tau}{2},$ which yields a critical point $k^* = \sqrt{2 l / \mu \tau}$ (only the positive value is feasible). Further, the second derivative yields $\frac{\partial^2 T_{l}(k)}{\partial k^2} = \frac{2 l}{\mu k^3} > 0 \quad \forall k \in (1,K]$. Thus, $T_l(k)$ is convex for real $k$ and achieves its minimum at $K_l^* = [k^*]$.
\EndProof

\noindent \textbf{Remarks.} 2. Lemma \ref{lem:serve-set} shows that the number of queues that will serve population $l$, $K_l$, is proportional to $\sqrt{l}$, when $\mu \tau < 1$. \\
% Note that, if $\mu \tau \geq 1$, there is at most one queue that serves the first arriving population. On the other hand if $\mu \tau \geq 2$, then the number of queues serving populations 1 and 2 will be at most one, again. This process can be iterated for each arriving population to determine if there is at most one queue that serves that population.
3. If $(l-1) \leq \mu \tau < l$, then, there will be more than one queue that serves population $l$, and at most one that serves all populations with index less than $l$. In this case as well, the number of queues that serve population $l$ can be found by solving a convex optimization problem. 

Now, we characterize the price of anarchy in the special case when every queue in the network serves at the same rate, and service start times are equi-spaces.

\begin{theorem} \label{theorem:case2PoA}
Suppose that each queue offers the same service rate $\mu > 0$, and queue $k$ starts service at time $\tau (k-1)$ with $\t > 0$. Then, the price of anarchy $\eta$ is
\[
\eta = \frac{  \frac{\tau}{2} \{  (1 - \frac{1}{\gamma_N})\sum_{l=1}^N \alpha_l + \sum_{l=1}^N \beta_l \} }
{ \frac{\mu}{2}  \sum_{l=1}^N \beta_l \bigg \{ \frac{\tau^2}{12}\sqrt{\frac{2}{\mu \tau}} (\sqrt{l} - \sqrt{l-1}) + \frac{2 \tau^2}{3} (\frac{2}{\mu \tau})^{3/2} (l \sqrt{l} - (l-1) \sqrt{l-1}) - \frac{\tau}{\mu} \bigg \} } \leq 2.
\]
\end{theorem}
\Proof
  The expression for $\eta$ follows by substituting, for any queue $l = 1, \cdots, K$, $\mu_{l} = \mu$ and $T_{s,l} = \tau (l-1)$ (i.e., the queue $l$ starts service at time $\tau (l-1)$), in \eqref{social-cost-eq} and \eqref{social-cost-opt}, using Lemma \ref{lem:serve-set} and taking $K_l \approx \sqrt{2 l / \mu \tau}$. The expression in the statement then follows after some elementary algebra and is omitted for brevity. 

To see that $\eta$ is upper-bounded by 2, first consider $N = 3$. The expression for $J_{opt}$ reduces to
\begin{eqnarray*}
J_{opt} = &\beta_1& \frac{\tau}{2} \bigg ( \frac{4}{3} \sqrt{\frac{2}{\mu \tau}} + \frac{\mu \tau}{12} \sqrt{\frac{2}{\mu \tau}} - 1\bigg ) + \frac{\mu}{2} \beta_2 \bigg ( \tau \sqrt{\frac{2 \tau}{\mu}} \bigg ( \frac{\sqrt{2} - 1}{12} + \frac{4}{3 \mu \tau} (2 \sqrt{2} - 1)\bigg ) -\frac{\tau}{\mu} \bigg ) \\ &\:+& \frac{\mu}{2} \beta_3 \bigg ( \tau \sqrt{\frac{2 \tau}{\mu}} \bigg ( \frac{\sqrt{3} - \sqrt{2}}{12}  + \frac{4}{3 \mu \tau} (3 \sqrt{3} - 2 \sqrt{2})\bigg ) - \frac{\tau}{\mu} \bigg ).
\end{eqnarray*}
$J_{eq}$ is simply $\frac{\tau}{2} \bigg (1-\frac{1}{\gamma_3} \bigg) \sum_{i=1}^3 \alpha_{\pi(i)} + \frac{\tau}{2} \sum_{i=1}^3 \beta_{\pi(i)} \equiv \frac{\tau}{2} \bigg (1-\frac{1}{\gamma_3} \bigg) \sum_{i=1}^3 \alpha_{i} + \frac{\tau}{2} \sum_{i=1}^3 \beta_{i}$. Using these expressions, we evaluate $2 J_{opt} - J_{eq} = $
\begin{eqnarray*}
\frac{\tau}{2} \sum_{i=1}^3(\alpha_i) \bigg ( \frac{1}{\gamma_3} - 1 \bigg ) &+& \beta_1 \frac{\tau}{2} \bigg ( \frac{4}{3} \sqrt{\frac{2}{\mu \tau}} + \frac{\mu \tau}{12} \sqrt{\frac{2}{\mu \tau}} - \frac{3}{2} \bigg ) 
+ \beta_2 \bigg (  \mu \tau \sqrt{\frac{2 \tau}{\mu}} \bigg ( \frac{\sqrt{2} - 1}{12} + \frac{4}{3 \mu \tau} (2 \sqrt{2} - 1) \bigg ) - \frac{3 \tau}{2} \bigg )\\ 
 &\:+& \beta_3 \bigg ( \mu \tau \sqrt{\frac{2 \tau}{\mu}} \bigg ( \frac{\sqrt{3} - \sqrt{2}}{12} + \frac{4}{3 \mu \tau} (3 \sqrt{3} - 2 \sqrt{2}) \bigg ) - \frac{3 \tau}{2} \bigg ).
\end{eqnarray*}
The first term on the right hand side is $> 0$, since $\gamma_3 < 1$ and $\alpha_i \geq 0$ for all $i$. The terms after $\beta_2$ and $\beta_3$ can easily be verified to be non-negative. The only term left to consider is the one after $\beta_1$. Denote $\delta = \frac{4}{3} \sqrt{\frac{2}{\mu \tau}} + \frac{\mu \tau}{12} \sqrt{\frac{2}{\mu \tau}} - \frac{3}{2}$. Multiplying and dividing by $\sqrt{2/\mu \tau}$ we have $\delta = \sqrt{\frac{\mu \tau}{2}} \left ( \frac{4}{3} \frac{2}{\mu \tau} + \frac{1}{6} - \frac{3}{2} \sqrt{\frac{2}{\mu \tau}} \right )$. Let $x := \sqrt{2/\mu \tau}$. Then, it can be seen that $\delta x = (\frac{4}{3}x^2 - \frac{3}{2} x + \frac{1}{6})$. Suppose that $\d x < 0$. That is, (after factoring the LHS) $(8x - 1) (x-1) < 0$. This implies either $(8x -1) > 0$ and $(x - 1) < 0$, which contradicts the fact that $x = \sqrt{2/\mu \tau} > 1$ when $\mu \tau < 1$; or $(8 x - 1) < 0$ and $(x - 1) > 0$ which is impossible. Therefore, it cannot be the case that $(8x - 1) (x-1) < 0$, thus proving that $2 \geq J_{eq} / J_{opt}$. It can also be checked (after some tedious algebra) that the terms after $\beta_l$, for $l > 3$, are larger than those after $\beta_3$ and so it is possible to use the same argument for an arbitrary number of arriving populations, $N$.
\EndProof

%----------------------------------------------------------
\section{Exact equilibrium analysis: An example with two users} 
\label{sec:finite}

A natural question is whether equilibria can be found in a finite population strategic arrival game, and how close the non-atomic equilibrium is to such an outcome.  It turns out that finite population equilibrium analysis is not malleable to a tractable analysis in general, and the resulting expressions can be solved exactly only in some special cases. We illustrate the methodology with two FIFO queues serving in parallel, with two arriving users, with service rates $\mu_k$, service start times $T_{s,1} = T_{s,2} = 0$. We derive the exact equilibrium arrival distribution, and show that it is unique. The extension to more than two users cannot be solved in closed form. Finite population analysis for a single server queue can be found in \cite{JuSh2011}, which also makes the same observation.

Let the two users be indexed by $i \in \{1,2\}$, and let $T_i$ be the arrival time of user $i$. Assume that the distribution function $F_i$ of $T_i$ is absolutely continuous. Let $\xi_i (t) = k \in \{1,2\}$  be a routing random variable, such that $\xi_i(\omega,t) = k$ implies that user $i$ at time $t$ is routed to queue $k$.  Let $p_{i,k}(t) := P(\xi_i(t) = k)$ be the probability of user $i$ being routed to queue $k$ at time $t$.  Let $Q^i_k (t)$ be the number of other arrivals that user $i$ observes in queue $k$ at time $t$, and let $A^i_k (t)$ denote the aggregate number of arrivals, other than the user $i$, to queue $k$ queue by time $t$, and let $S_k (t)$ denote the cumulative number of (potential) service completions, at queue $k$, up to time $t$. 

As noted before, the path-wise description of the queue length process is given by $Q^i_k(t) = (A^i_k(t) - S_k(t))_+, \quad t \in \mathbb{R}, \quad i \in \{1,2\} ~\text {and}~ k \in \{1,2\}$. We also make the assumption that the routing, arrival and service processes are mutually independent. We denote by $-i$, the user $j \neq i$. The following proposition describes the dynamics of the expected queue length, $\bar{Q}^i_k(t)$.

\begin{proposition} \label{prop:expected-queue-length}
% Let $\bar{Q}^i_j(t)$ be the expected queue length at time $t \in \mathbb{R}$, which evolves as follows:
\(
\bar{Q}^i_k(t+dt) = \bar{Q}^i_k(t) + p_{k}(t) f_{ -i }(t) dt - \mu_k (1 - P_{k,0}(t)) dt \mathbf{1}_{\{t \geq 0\}} \quad \forall i  \in \{1,2\} \text{ and } k \in \{1,2\},
\)
where, $f_{-i}(\cdot)$ is the density function of the arrival time of the user other than user $i$ and $P_{k,0}(t)$ is the probability that the queue $k$ is empty at time $t$, i.e., $P_{k,0}(t) = P(Q^i_k(t) = 0)$. 
\end{proposition}
\Proof
It is easy to see that $\bar{Q}^i_k(t+dt) = \bar{Q}^i_k(t) + \bbE[A^i_k(t+dt) - A^i_k(t)] - \bbE[D_k(t+dt) - D_k(t)] \mathbf{1}_{\{t \geq 0\}}$. Here, $D_k(t)$ is the number of actual service completions by time $t$ and $\bbE[A^i_k(t+dt) - A^i_k(t)]$ is the expected number of users other than $i$ who might arrive in the interval $[t,t+dt]$.  The expectation of user $-i$ arriving in an infinitesimal time interval is $p_{k}(t) f_{-i}(t) dt$. The expected number of service completions in the interval $[t,t+dt]$ is given simply by $\mu_k(1- P(Q^i_k(t) = 0))dt$. The expression for $\bar{Q}^i_k(t)$ follows by substitution.
\EndProof
The mean virtual waiting time of user $i$ at queue $k$ at time $t$ is given by $\bar{W}^i_k(t) = \frac{\bar{Q}_k^i(t)}{\mu_k} - t \mathbf{1}_{\{t \leq 0\}}$% , i.e., the expected workload seen by a (potential) arrival at time $t$ is dependent upon the expected amount of time required to serve out the users in the system at time $t$, and whether the arrival is before or after service commences
. Note that an arrival at time $t < 0$, would have to wait for $t$ units of time before service commences, explaining the presence of the term $t \mathbf{1}_{\{t \leq 0\}}$.

% We now describe the strategic arrival game in this two user setting.
Now, users choose a mixed strategy over the arrival interval and the routing random variable. Thus, the strategy space of the game is $\mathcal{C}_b \times [0,1]^2$, where $\mathcal{C}_b$ is the space of non-decreasing and absolutely continuous functions with support in $\mathbb{R}$, and $[0,1]^2$ is the set that the routing probabilities take values in. Thus, user $i$ chooses the tuple $(F_i, \mathbf{p}_i)(t), ~\forall t \in \mathbb{R}$, where $\mathbf{p}_i(t) = (p_{1,i}(t), p_{2,i}(t))^{'}$ is a vector of probabilities at time $t$. Thus, the choice of $\mathbf{p}_i(t)$ must satisfy the constraint $p_{1,i}(t) + p_{2,i}(t) = 1$. We designate $\textbf{F} = (F_1,F_2)$ as an arrival profile and $\textbf{P} = (\mathbf{p}_1, \mathbf{p}_2)$ as a routing profile of the game. We are interested in symmetric equilibria in this one-shot arrival game, such that $F_1=F_2$ and $\mathbf{p}_1 = \mathbf{p}_2$. For brevity, we drop the superscripts from $Q_k^i$ and $A_k^i$, as the definition should be clear from the context.

Now, as before, the expected cost of arriving at queue $k$ is a weighted sum of the waiting time and the time at which the user arrives, which yields $C_k(t) = (\alpha + \beta) \left ( \frac{\bar{Q}_k(t)}{\mu_k} - t \mathbf{1}_{\{t \leq 0\}} \right ) + \beta t$. Note that the equilibrium arrival profile will have some finite support $[-T_{0,k},T_k]$ for queue $k$ since the expected cost is increasing in both $t$ and $-t$ (since there is only one other user, the expected queue length is bounded by 1). Furthermore, the expected cost of arrival must be the same at either queue. If this were not the case, then an arriving user could improve its cost by choosing to arrive at a queue with a  lower cost. We are interested in symmetric equilibria for which the arrival interval must be the same at either queue. Let this interval be $[-T_0,T]$, where $-T_0$ and $T$ are determined in equilibrium.

%Let the equilibrium arrival profile be $\textbf{F}^*$ and the equilibrium routing profile be $\textbf{P}^*$. The following theorem characterizes the equilibrium profiles.

\begin{theorem} \label{thm:eq-profile-finite}
The equilibrium profile, ($\textbf{F}^*,\textbf{P}^*$), of the strategic arrivals game with two users and two queues is given by
\[
f^*(t) = \begin{cases} & \gamma (\mu_1 + \mu_2) \quad t \in [-T_0,0]\\ & \gamma(\mu_1 + \mu_2) - \mu_1 P_{1,0}(t) - \mu_2 P_{2,0}(t) \quad t \in (0,T], \end{cases}
\]
\[
\quad\text{and}~~~ p^*_i(t) \,=\, \begin{cases} & \frac{\mu_i}{(\mu_1 + \mu_2)}, \quad t \in [-T_0,0] \\ 
& \frac{\mu_i}{(\mu_1 + \mu_2)} + (\mathbf{1}_{\{i=1\}} - \mathbf{1}_{\{i = 2\}}) \frac{\mu_1 \mu_2 (\mu_2 - \mu_1)}{(\mu_1 + \mu_2) f(t)} \frac{\beta t}{(\alpha + \beta)},  \quad t \in (0,T],\end{cases}
\]
where $\g = \a/(\a+\b)$, $P_{i,0}(t) = 1 - \frac{\m_i(\b t - \a T_0)}{\a + \b}$ and
\[
T_0 = - \bigg ( \frac{\mu_1 + \mu_2}{(\mu_1^2 + \mu_2^2)} \bigg ) \sqrt{\bigg (2 + \frac{\beta}{\alpha} \bigg)\frac{\beta}{\alpha}} \quad
\text{ and } \quad
T = \bigg ( \frac{\mu_1 + \mu_2}{\mu_1^2 + \mu_2^2} \bigg ) \bigg ( \sqrt{\frac{2 \alpha}{\beta} + 1} - 1 \bigg ).
\]
\end{theorem}
The proof can be found in the appendix.

\noindent \textbf{Remarks.} % 1. It should be noted that the expressions for $T$ and $-T_0$ are  generalizations of those for the single server queue, derived in \cite{JuSh2011}. This is intuitively satisfying, since the two queues start service at the same instant. However, the arriving users to the parallel queue network have to strategize over the routing probabilities as well.\\
1. Note that the general case of $K$ parallel queues serving $n$ users is not as simple as the result above. The network state dynamics are determined by a set of $K$ coupled differential equations, that are in general quite difficult to solve. Instead, using a fluid approximation reduces the complexity of the problem, by allowing one to replace the differential equations by simple linear equations. 

% Thus, the equilibrium in the fluid limit derived earlier affords us considerable simplification since the path-wise description of the system dynamics are reduced to tractable deterministic equations, and otherwise it is a hopelessly complicated fixed point problem. 
2. The fluid analysis does lose some accuracy, however, for small numbers of arrivals since it suggests that queues can never be idle at equilibrium, but for a large number of users it still captures the \emph{essential} features of the strategic arrivals game.

\section{Conclusions} \label{sec:conclusion}
In this paper, we have presented three results. First, we have developed large population fluid approximations to various processes of interest in a parallel queueing network where the arriving users choose a time of arrival from an arbitrary distribution function. We believe these are entirely new results and should be of independent interest. Second, using this framework, we then studied the \textit{network concert queueing game} in the large population regime. We proved the existence and uniqueness of the non-atomic equilibrium arrival profile, both in the case of a homogeneous population of users, as well as heterogeneous populations with disparate cost characteristics. In either case, we also characterized the \textit{price of anarchy} of the game, due to strategic arrivals. Third, we demonstrated the methodology for finite population analysis, by analyzing a simple instance with two strategic users arriving at a two queue network, and deriving the equilibrium arrival profile. 

%There are a number of situations where users arrive at a network of parallel queues, while strategizing their arrival time. For instance, consider passengers arriving at an airport for a flight. The arriving users face a number of ``security lanes'' operating in parallel at the security checkpoint. The passengers arriving at the security checkpoint can choose their time of arrival and the routing probability strategically, so as to minimize some appropriate cost function. Similarly, one can consider the case of computer users, or tasks, being queued up at cloud servers that assign a virtual machine to each user. If the cost function used is the delay experienced by a user, then, users will strategically choose a time of submission and an assignment to one of the servers, according to the equilibrium outcome, to reduce the waiting time for the completion of the user.

% As noted in the introduction, the practical applications of the results we have developed are manifest. 

The queueing model that we have introduced in this paper is of relevance in several settings including transportation networks, data center network traffic, etc. Thus, it would be useful to understand further the various stochastic processes. For example, the FSLLN/fluid limit analysis shows that at equilibrium, no queue is ever idle. This, however, cannot be true in the case of a finite population, where, as we have seen in Section \ref{sec:finite}, there is a positive probability of the queue being idle during the arrival interval. Thus, we must derive better approximations to the arrival, queue-length and waiting-time processes. One way would be to derive diffusion limits by developing functional central limit theorems (FCLT) for the queueing system model we introduced. This is a current line of our research. In particular, we have been able to establish that the diffusion limiting process to the queue-length in a single server queue is combination of a Brownian motion and a Brownian bridge, where the weak convergence has been established in Skorokhod's $M_1$ topology.

Finally, the alternative queueing system model we have introduced in this paper may involve a general queueing network. In such a setting, it would be useful to develop the fluid and diffusion limits for the various processes of interest, and to establish the set of (non-atomic) strategic equilibria that result. We are currently working towards this goal. Our hope is that the framework that we have introduced would be of greater relevance for some scenarios, as well as more tractable than the GI/G/1 models in standard queueing theory. 

\section*{Appendix}

%-----------------------------------------------
We first state two Lemmas that are useful in proving Theorem \ref{thm:fluid-primitives}, and are also useful below.

\begin{lemma} \label{lem:arrival-fluid}
Let $A^n_k(t), \,\, \forall k$, be as defined in \eqref{arrival-fluid-scale}, where $n$ is the population size, and $\mathbf{p} = (p_{0,1}, \cdots, p_{0,K})^{'}$. Then, $\bar{A}^n(t) := \frac{A^n(t)}{n}  \rightarrow F(t) \quad a.s.~u.o.c.~\forall t \in [-T_0,T]$, as $n \rightarrow \infty$.
% \begin{eqnarray}
% \label{cumulative-arrival-fluid}
% \bar{A}^n(t) := \frac{A^n(t)}{n}  &\rightarrow& F(t) \quad a.s. \quad u.o.c., \,\, \forall t \in [-T_0,T], \text{ and }\\
% \label{arrival-vector-fluid}
% \frac{1}{n} \mathbf{A}^n(t)  &\rightarrow& \bar{\mathbf{A}}(t) := \mathbf{p}
% F(t) \quad a.s. \quad u.o.c., \,\, \forall t \in [-T_0,T],
% \end{eqnarray}
\end{lemma}

\begin{lemma} \label{lem:service-fluid}
Let $\bigg (\frac{1}{n}S_k^n(t), V_k^n(t+T_0) \bigg )$ be as defined in \eqref{service-fluid-scale} and \eqref{service-time-fluid-scale}, where $k \in \{1,\cdots,K\}$ and $t \in [-T_0,\infty)$. Then, as $n \rightarrow \infty$,
\begin{eqnarray*}
\bigg (\frac{1}{n} \mathbf{S}^n(t), \mathbf{V}^n(t+T_0) \bigg )
\rightarrow (\bar{\mathbf{S}}(t), \bar{\mathbf{V}}(t)) :=
(\mathbf{M}^{-1} \mathbf{t_{s,+}}(t), \mathbf{M} \mathbf{1} (t+T_0)) \quad a.s.~u.o.c.~\quad \forall \, t \in [-T_0, \infty),
\end{eqnarray*}
where $M :=
diag(m_1, \cdots, m_K)$ is a diagonal matrix of the mean service times at
each node, $\mathbf{t_{s,+}}(t) := ((t-T_{s,1}))\mathbf{1}_{\{t \geq T_{s,1}\}}, \cdots, (t-T_{s,K}) \mathbf{1}_{\{t
  \geq T_{s,K}\}})^{'} $ and $\mathbf{1} = (1,\cdots,1)^{'}$. 
\end{lemma}

Both Lemmas above can be proved easily using the Strong Law of Large Numbers, and their proofs are omitted due to space constraints.

%---------------------------------------
\subsection*{Proof of Proposition \ref{prop:X-fluid}}

\Proof (Proposition \ref{prop:X-fluid})
In order to establish a functional strong law of large numbers
result for \eqref{X-fluid-scale} we first require the following lemma
relating the processes $\mathbf{I}_k^n(t) := \bigg ( \int^t_{T_{s,k}} \mathbf{1}_{\{Q_k^n(s) = 0\}} ds\bigg ) \mathbf{1}_{\{t \geq T_{s,k}\}}$ and $\tilde{\mathbf{I}}^n(t) := \int^t_{-T_0} \mathbf{1}_{\{Q_k^n(s) = 0\}} ds$.
\begin{lemma}
\label{lem:idle}
Let $I_k^n(t)$ and $\tilde{I}_k^n(t)$, the $k$th components of
$\mathbf{I}^n(t)$ and $\tilde{\mathbf{I}}^n(t)$, be elements of
$\mathcal{D}$. Then, as $n \rightarrow \infty$,
\(
| I_k^n(t) - \tilde{I}_k^n(t) | \rightarrow 0 \,\,a.s. \quad u.o.c.\,\,
\forall t \in [-T_0,\infty).
\)
Thus, $|\mathbf{I}^n(t) -\tilde{\mathbf{I}}^n(t) | \rightarrow 0, \,\,a.s. \quad u.o.c.\,\,
\forall t \in [-T_0,\infty)$.
\end{lemma}
The proof of Lemma \ref{lem:idle} follows directly from the following result.

\begin{lemma}
\label{lem:first-arrival-time}
Let $(\Omega,\mathcal{F},P)$ be a probability space. $F(t)$ is a
distribution function of a random variable $\mathcal{T}$ on this space, with support
$[S,T] \, \subset \mathbb{R}$. Assume that $F(t)$ is absolutely continuous. Suppose
$T_1, T_2, \cdots, T_n$ are $n$ I.I.D. samples drawn from
$\mathcal{T}$. Define $T_1^{(n)} := \inf
\{ T_1, T_2, \cdots \, T_n\}$ as the value of the smallest sample. Then, it follows that,
\(
T_1^{(n)} \downarrow S \quad a.s. \quad \text{ as } n \rightarrow \infty.
\)
\end{lemma}
\Proof
We will prove this in the case of a uniformly distributed
random variable, on $[0,1]$. Let $U(t)$ be the distribution function
of the standard uniform random variable, and let $F(t) = U(t)$. The general case then follows by making use of a
transformation of the uniform random variable, since $F(t)$ is assumed
to be absolutely continuous. Fix $\epsilon > 0$. Consider the event,
\(
E_n \,:=\, \{\omega: T_1^{(n)} \geq \epsilon \}.
\)
The measure of this event is given by,
\(
P(E_n) \,=\, P(T_1^{(n)} \geq \epsilon) \,=\, P(\inf \{ T_1, \cdots,
T_n\} \geq \epsilon).
\)
Since the samples are I.I.D., it follows easily that,
\(
P(E_n) = [P(T_1 \geq \epsilon)]^n = [1 - F(\epsilon)]^n.
\)
Since $F(t)$ is absolutely continuous, it follows that,
\(
\sum_{n=1}^{\infty} P(E_n) \,<\, \infty.
\)
By the First Borel- Cantelli Lemma, it follows that $P(T_1^{(n)} \geq
\epsilon \,\,\, i.o.) = 0$. This implies that $T_1^{(n)} \downarrow S = 0
\,\, a.s.$. 
\EndProof
\Proof (Lemma \ref{lem:idle})
Let $\tau > 0$. As the queue is non empty after the very first arrival to the node, we have $I_k^n(t) \equiv (\int_{T_{k,1}^{(n)}}^{t} \mathbf{1}_{\{Q_k^n(s) = 0\}} ds) \mathbf{1}_{\{t \geq T_{s,k}\}}$. It is easy to see that $I_k^n(t) \leq \tilde{I}_k^n(t) \quad \forall t \in
[-T_0,\tau]$. Also, $I_k^n(t) \geq \tilde{I}_k^n(t) -
\int_{-T_0}^{T_{k,1}^{(n)}} \mathbf{1}_{\{Q_k^n(s) = 0\}} ds, \quad \forall t
\in [-T_0,\tau]$. It follows that
\(
\int_{-T_0}^{T_{k,1}^{(n)}} \mathbf{1}_{\{Q_k^n(s) = 0\}} ds \,\geq \,
\tilde{I}_k^n(t) - I_k^n(t) \,\geq\, 0.
\)

The conclusion follows on $[-T_0,\tau]$, by noting that $Q_k^n(s) = 0 \,\, \forall s
\in [-T_0,T^{(n)}_{k,1}]$ and applying Lemma
\ref{lem:first-arrival-time}. Since $\tau$ is arbitrary, the lemma is
proved.
\EndProof
Now, we can proceed to prove Lemma \ref{prop:X-fluid}, and we analyze the claim component-wise, since the processes at each node are statistically independent. The result follows by applying Lemmas \ref{lem:arrival-fluid}, \ref{lem:service-fluid} and \ref{lem:idle} to \eqref{X-fluid-scale}. First, note that $B^n_k(t) \leq t$. Thus, by the Random Time Change Theorem (Theorem 5.5, \cite{ChYa2001}) and Lemma \ref{lem:service-fluid} it follows that
\(
\frac{1}{n}S_k^n(B_k^n(t)) - \mu_k B_k^n(t) \rightarrow 0,
\)
$\,\, a.s. \quad u.o.c. \text{ as } n \rightarrow \infty \,\, \forall
t \in [-T_0,\infty)$. By Lemma \ref{lem:arrival-fluid} we have
\(
\frac{1}{n} A_k^n(t) - p_{0,k} F(t) \rightarrow 0,
\)
$ \,\, a.s. \quad u.o.c. \text{ as } n \rightarrow \infty \,\, \forall
t \in [-T_0,\infty)$. Applying these results and Lemma \ref{lem:idle}
to \eqref{X-fluid-scale} we have
$X_k^n(t) \rightarrow \bar{X}_k(t) := (p_{0,k} F(t) - \mu_k
(t-T_{s,k})\mathbf{1}_{\{t \geq T_{s,k}\}})\,\, a.s. \quad u.o.c. \,\, \text{ as
} n \rightarrow \infty$. It follows that
\(
\mathbf{X}^n(t) \rightarrow (\mathbf{p} F(t) -
\mathbf{M}^{-1} \mathbf{t_{s,+}}(t)),
\)
$\,\, a.s. \quad u.o.c., \text{ as } n \rightarrow \infty, \,\, \forall
t \in [-T_0,\infty)$.
\let\IEEEQED\relax
\EndProof

%-----------------------------------------------
\subsection*{Proof of Theorem \ref{thm:queue-fluid}}
\Proof
Recall that $\frac{1}{n}\mathbf{Q}^n(t) := \mathbf{X}^n(t) + \mathbf{Y}^n(t)$, where $\mathbf{X}^n$ and $\mathbf{Y}^n$ are defined in \eqref{X-fluid-scale} and \eqref{Y-fluid-scale}, respectively.
%( \frac{1}{n} \mathbf{A}^n(t) -
%\mathbf{p} F(t) \bigg ) - \bigg ( \frac{1}{n} \mathbf{S}^n(B^n(t))
%- \mathbf{M}^{-1} \mathbf{B}^n(t)\bigg ) + (\mathbf{p} F(t) -
%\mathbf{M}^{-1} \mathbf{t_{s,+}}(t))+ \mathbf{M}^{-1} (\mathbf{I}^n(t) -
%\tilde{\mathbf{I}}^n(t)), \text{ and } \mathbf{Y}^n(t) := \mathbf{M}^{-1} %\tilde{\mathbf{I}}^n(t)$.
Note that if we defined $\mathbf{Y}^n(t) := \mathbf{I}^n(t) = \mathbf{t}_{s,+}(t) - \mathbf{B}^n(t)$, the Oblique Reflection Mapping Theorem (Theorem 7.2, \cite{ChYa2001}) would not be satisfied, since this process could be zero even when $\mathbf{Q}^n(t)$ is zero. To see this, note that if $T_1^{(n)}$ is the time of the first arrival to the network then, $\forall t \in
[-T_0,T_1^{(n)}]$, $\mathbf{I}^n(t)$ and $\mathbf{Q}^n(t)$ would be
zero, violating the Oblique Reflection Mapping Theorem.

% The following lemma shows that the queue length process of the
% parallel node network satisfies the
% Oblique Reflection Mapping with $P = 0$. 
% \begin{lemma}
% \label{lem:queue-ORT}
% Let $\frac{1}{n}\mathbf{Q}^n$, $\mathbf{X}^n$ and $\mathbf{Y}^n$, as
% defined in \eqref{vector-queue-length-fluid-scale}
% \eqref{X-fluid-scale} and \eqref{Y-fluid-scale} respectively, be
% elements of $\mathcal{D}^K$. Then, 
% \(
% \bigg ( \frac{1}{n}\mathbf{Q}^n,\mathbf{Y}^n \bigg ) = (\Phi(\mathbf{X}^n),\Psi(\mathbf{X}^n)).
% \)
% \end{lemma}
% \Proof
It is a simple exercise to verify that Theorem 7.2 of \cite{ChYa2001} is satisfied, in the case of a parallel node network, with $\mathbf{Y}^n(t) = \mathbf{M}^{-1} \tilde{\mathbf{I}}^n(t)$: The zero matrix $P$ is trivially a $M$-matrix. By
definition, we have $\mathbf{Q}^n(t) \geq 0, \,\, \forall t \in
[-T_0,\infty)$. $\mathbf{Y}^n(t)$ is a non-decreasing $K$-dimensional
process that only grows when every component of $\mathbf{Q}^n(t)$ is zero.%  Thus, the conditions of Theorem 7.2 of \cite{ChYa2001} are satisfied.
%\EndProof

% We can now easily prove Theorem \ref{thm:queue-fluid}. It follows by Lemma \ref{lem:queue-ORT} that $(\mathbf{Q}^n(t)/n,
% \mathbf{Y}^n(t))$ satisfy Theorem Theorem 7.2, \cite{ChYa2001}. 
This implies,
\(
( \frac{1}{n}\mathbf{Q}^n(t), \mathbf{Y}^n(t) ) = (\Phi(\mathbf{X}^n(t)), \Psi(\mathbf{X}^n(t))).
\)
The reflection regulator map, $\Psi(\cdot)$, is Lipschitz continuous under the uniform metric (Theorem 7.2, \cite{ChYa2001}). By the Continuous Mapping Theorem and Lemma \ref{prop:X-fluid} it follows that
\(
(\Phi(\mathbf{X}^n(t)), \Psi(\mathbf{X}^n(t))) \rightarrow (\Phi(\bar{\mathbf{X}}(t)), \Psi(\bar{\mathbf{X}}(t))),  \,\, a.s. \quad u.o.c. \text{ as } n \rightarrow \infty, \,\, \forall t \in [-T_0,\infty).
\)
\let\IEEEQED\relax
\EndProof

%--------------------------------------------------
\subsection*{Proof of Theorem \ref{thm:fluid-busy-virtual}}

\Proof We prove the Theorem by treating the busy time and virtual waiting time processes separately. The fluid limit of the busy time process is derived in the following Lemma.
\begin{lemma}
\label{theorem:busy-time-fluid}
Let $\mathbf{B}^n := (B_1^n, \cdots, B_K^n)^{'}$ be an
element of $\mathcal{D}^K$. Then as $n \rightarrow \infty$
\(
\mathbf{B}^n(t) \rightarrow \bar{\mathbf{B}}(t) := \mathbf{t_{s,+}}(t) - \mathbf{M}
\Psi(\bar{\mathbf{X}}(t))
 = \mathbf{t_{s,+}}(t) - \mathbf{M} \sup_{-T_0 \leq s
  \leq t} [-\bar{\mathbf{X}}(s)]_+ \,\, u.o.c. \,\, a.s.,\,\, \forall t \in [-T_0,\infty),
\)
where,
$
\mathbf{t_{s,+}}(t) = (t \mathbf{1}_{\{t \geq T_{s,1}\}}, \cdots, t \mathbf{1}_{\{t \geq
  T_{s,K}\}})^{'}
$
and
$
M = diag(m_1, \cdots, m_K).
$
\end{lemma}
\Proof
By definition, we have, 
\(
\mathbf{B}^n(t) = \mathbf{t_{s,+}}(t) - \mathbf{I}^n(t).
\)
Adding and subtracting the vector process $\tilde{\mathbf{I}}^n(t)$
and noting that $\mathbf{Y}^n(t) = \mathbf{M}^{-1}
\tilde{\mathbf{I}}^n(t)$ we have
\(
\mathbf{B}^n(t) = \mathbf{t_{s,+}}(t) - \mathbf{M} \mathbf{Y}^n(t) + \tilde{\mathbf{I}}^n(t) - \mathbf{I}^n(t).
\)
Using Lemma \ref{lem:idle} from the Appendix and Theorem \ref{thm:queue-fluid} it follows that as $n \rightarrow \infty$,
\(
\mathbf{B}^n(t) \rightarrow \mathbf{t_{s,+}}(t) - \mathbf{M} \Psi(\bar{\mathbf{X}}(t)) \,\, a.s. \quad u.o.c., \,\, \forall t \in [-T_0,\infty).
\)
\EndProof
Next, we derive the fluid limit of the virtual waiting time process in the following Lemma.
\begin{lemma}
\label{theorem:virtual-waiting-fluid}
Let $\mathbf{W}^n = (W_1^n, \cdots, W_K^n)^{'} \in \mathcal{D}^K$. Then as $n \rightarrow \infty$
\(
\mathbf{W}^n(t) \rightarrow \bar{\mathbf{W}}(t) := \mathbf{M}
\mathbf{p} F(t) - \bar{\mathbf{B}}(t) - \mathbf{t_{s,-}}(t) \quad a.s. \quad u.o.c.,\,\, \forall t \in [-T_0, \infty),
\)
where $\mathbf{t_{s,-}}(t) = ((t-T_{s,1})
\mathbf{1}_{\{t \leq T_{s,1},\}}, \cdots, (t-T_{s,K}) \mathbf{1}_{\{t \leq T_{s,K}\}})^{'}$, $\mathbf{p} = (p_{0,1}, \cdots, p_{0,K})^{'}$ and $\mathbf{M} =
diag(m_1, \cdots, m_K)$.
\end{lemma}
\Proof
The theorem follows by an application of Lemma
\ref{thm:fluid-primitives}, the Random Time Change Theorem (Theorem 5.5,
\cite{ChYa2001}) and Lemma \ref{thm:fluid-busy-virtual}. We define the following vector for notational convenience. Let
\(
\mathbf{V}(A(t)) := (V_1(A_1(t)), \cdots, V_K(A_K(t)))^{'}
\)
The fluid-scaled version of the virtual waiting time process is given by
\(
\mathbf{W}^n(t) = \mathbf{V}^n(A^n(t)) - \mathbf{B}^n(t) - \mathbf{t_{s,-}}(t).
\)
First note that by Lemma \ref{lem:service-fluid} and the Random Time
Change Theorem $\mathbf{V}^n(A^n(t)) -\mathbf{M} \frac{1}{n} \mathbf{A}^n(t) \rightarrow
0\,\, u.o.c. \quad a.s. \text{ as } n \rightarrow \infty, \,\, \forall t \in
[-T_0,\infty)$. Thus, centering the process $\mathbf{V}^n(A^n(t))$ we
get
\(
\mathbf{W}^n(t) = \mathbf{V}^n(A^n(t)) - \mathbf{M} \frac{1}{n}
\mathbf{A}^n(t) + \mathbf{M} \frac{1}{n} \mathbf{A}^n(t) - \mathbf{B}^n(t) -\mathbf{t_{s,-}}(t).
\)
Applying Lemma \ref{lem:arrival-fluid}, Lemma
\ref{thm:fluid-busy-virtual} and the comment above we have
\(
\mathbf{W}^n(t) \rightarrow \mathbf{M} \mathbf{p} F(t) -
\bar{\mathbf{B}}(t) - \mathbf{t_{s,-}}(t).
\)
Substituting for $\bar{\mathbf{B}}(t)$ we get
\(
\bar{\mathbf{W}}(t) = \mathbf{M} \mathbf{p} F(t) - (\mathbf{t_{s,+}}(t) -
\mathbf{M} \Psi(\bar{\mathbf{X}}(t))) - \mathbf{t_{s,-}}(t) = \mathbf{M} \bar{\mathbf{Q}}(t) - \mathbf{t_{s,-}}(t).
\)
\let\IEEEQED\relax
\EndProof

%-----------------------------------------------
\subsection*{Proof of Theorem \ref{thm:eq-profile-finite}}
\Proof
Recall that $C_i(t) = c$, a constant,  ~$\forall i \in \{1,2\}$. Noting that that expected queue length at time $-T_0$ is zero, it is easy to see that $c = \alpha T_0$. It follows that $C^{'}_i(t) = 0$. Thus, solving for $p_{i}(t)f(t)$ we obtain
\[
p_i(t) f(t) \,=\, \begin{cases} & \gamma \mu_i \quad t \in [-T_0, 0]\\ & \gamma \mu_i - \mu_i P_{i,0}(t) \quad t \in (0,T]. \end{cases}
\]
Using the fact that $p_{1}(t) + p_2(t) = 1$, we solve for $p_1(t)$ and $p_2(t)$ to obtain
\[
p_i(t) \,=\, \begin{cases} & \frac{\mu_i}{(\mu_1 + \mu_2)} \quad t \in [-T_0,0] \\ 
& \frac{\mu_i}{(\mu_1 + \mu+2)} + (\mathbf{1}_{\{i \equiv 1 \} } - \mathbf{1}_{\{i \equiv 2\}}) \frac{\mu_1 \mu_2 (\mu_2 - \mu_1)}{(\mu_1 + \mu_2) f(t)} \frac{\beta t}{(\alpha + \beta)} \quad t \in (0,T]. 
\end{cases}
\]
Thus, it can be seen that the probability of being routed to queue $i$ (in equilibrium) is constant before service commences, and is time dependent afterwards. The time dependence is determined by the expected idle time of the queue, as can be seen by solving for $f(t)$. Again, using the fact that $p_1(t) + p_2(t) = 1$, add the equations for $p_i(t) f(t)$, for $i \in \{1,2\}$, to obtain
\[
f(t) = \begin{cases} & \gamma (\mu_1 + \mu_2) \quad t \in [-T_0,0]\\ & \gamma(\mu_1 + \mu_2) - \mu_1 P_{1,0}(t) - \mu_2 P_{2,0}(t) \quad t \in (0,T]. \end{cases}
\]
Interestingly, the arrival distribution is piecewise continuous. It is uniform up to time $0$, at which point service commences, and is a continuous function of $t$ in $[0,T]$. Notice that when there are only two arriving users, the expected queue length observed by one of the arrivals is fully determined by the probability of idling. We have for queue $i$,
\(
\bar{Q}_i(t) = 1 \times P(Q_i(t) = 1) + 0 \times P(Q_i(t) = 0) = P_{i,1}(t).
\)
Now, using the fact that $P_{i,0}(t) + P_{i,1}(t) = 1$ and the fact that $C_i(t) = \alpha T_0$, we can solve for $P_{i,1}(t)$ to obtain
\(
P_{i,1}(t) = \frac{\mu_i(\beta t - \alpha T_0)}{(\alpha + \beta)} = 1 - P_{i,0}(t).
\)
Thus, we can now substitute for $f(t)$ on $(0,T]$ to obtain
\(
f(t) \,=\, (\gamma - 1)(\mu_1 + \mu_2) + \frac{\alpha T_0 - \beta t}{(\alpha + \beta)} (\mu_1^2 + \mu_2^2).
\)

By definition, we have $\int_{-T_0}^T f(t) dt = 1$. Note that $f(t)$ is a continuous function of $t$ in the interval $[0,T]$. By assumption, $F$ has no point masses. Thus, it follows that $f(T) = 0$. Now, using the fact that $\int_{-T_0}^T f(t) dt =1$ we have
\(
(\gamma -1) (\mu_1 + \mu_2) (T+T_0) - (\mu_1^2 + \mu_2^2) ( (1-\gamma) \frac{T^2 - T_0^2}{2} - \gamma T_0 (T+T_0) ) = 1.
\)
Next, using the fact that $f(T) = 0$, we have
\(
T = \frac{\mu_1 + \mu_2}{\mu_1^2 + \mu_2^2} - \frac{\alpha}{\beta} T_0.
\)
Substituting for $T$ in terms of $T_0$ from the expression above, and solving the resulting quadratic equation, we obtain
\[
T_0 = - \bigg ( \frac{\mu_1 + \mu_2}{(\mu_1^2 + \mu_2^2)} \bigg ) \sqrt{\bigg (2 + \frac{\beta}{\alpha} \bigg)\frac{\beta}{\alpha}} \quad \text{and} \quad T = \bigg ( \frac{\mu_1 + \mu_2}{\mu_1^2 + \mu_2^2} \bigg ) \bigg ( \sqrt{\frac{2 \alpha}{\beta} + 1} - 1 \bigg ).
\]
These expressions describe the symmetric equilibrium strategy $(F^*,\textbf{p}^*)$ in the case of two strategically arriving users and two parallel queues. The uniqueness of the equilibrium profile follows by construction. For given service rates and cost characteristics, it is clear that $T$ and $T_0$ are unique. It is easy to see that $T$ and $-T_0$ together fully determine $F^*$ and hence $\textbf{p}^*$. It follows that the arrival profile and routing profiles are unique.
\EndProof

%\let\IEEEQED\relax\EndProof

%-----------------------------------------------
\newpage
\bibliographystyle{../../style/ormsv080} 
\bibliography{refs-fluidarrivalgame}

\begin{thebibliography}{15}
\expandafter\ifx\csname natexlab\endcsname\relax\def\natexlab#1{#1}\fi
\expandafter\ifx\csname url\endcsname\relax
  \def\url#1{{\tt #1}}\fi
\expandafter\ifx\csname urlprefix\endcsname\relax\def\urlprefix{URL }\fi
\expandafter\ifx\csname urlstyle\endcsname\relax
  \expandafter\ifx\csname doi\endcsname\relax
  \def\doi#1{doi:\discretionary{}{}{}#1}\fi \else
  \expandafter\ifx\csname doi\endcsname\relax
  \def\doi{doi:\discretionary{}{}{}\begingroup \urlstyle{rm}\Url}\fi \fi

\bibitem[{Adlakha and Johari(2010)}]{AdJo10}
Adlakha, S., R.~Johari. 2010.
\newblock Mean field equilibrium in dynamic games with strategic
  complementarities.
\newblock {\it Submitted\/} .

\bibitem[{Billingsley(1968)}]{Bi1968}
Billingsley, P. 1968.
\newblock {\it {Convergence of Probability Measures}\/}.
\newblock Wiley \& Sons.

\bibitem[{Chen and Yao(2001)}]{ChYa2001}
Chen, H., D.D. Yao. 2001.
\newblock {\it {Fundamentals of Queueing Networks: Performance, asymptotics,
  and optimization}\/}.
\newblock Springer.

\bibitem[{Dube and Jain(2011)}]{DuJa11}
Dube, P., R.~Jain. 2011.
\newblock Bertrand equilibria and efficiency in markets for congestible network
  services.
\newblock {\it Submitted to Automatica\/} .

\bibitem[{Durrett(2010)}]{Du2010}
Durrett, R. 2010.
\newblock {\it {Probability: Theory and Examples, 4th Ed.}\/}.
\newblock Cambridge University Press.

\bibitem[{Glazer and Hassin(1983)}]{GlHa1983}
Glazer, A., R.~Hassin. 1983.
\newblock {?/M/1: On the Equilibrium Distribution of Customer Arrivals}.
\newblock {\it European Journal of Operational Research\/} .

\bibitem[{Hassin and Haviv(2003)}]{HaHa2003}
Hassin, R., M.~Haviv. 2003.
\newblock {\it {To Queue or not to Queue}\/}.
\newblock Kluwer Academic Publishers.

\bibitem[{Jain et~al.(2011)Jain, Juneja, and Shimkin}]{JaJuSh2010a}
Jain, R., S.~Juneja, N.~Shimkin. 2011.
\newblock {The Concert Queueing Game: To Wait or To be Late}.
\newblock {\it Discrete Event Dynamic Systems\/} {\bf 21(1)} 103--134.

\bibitem[{Juneja and Jain(2009)}]{JuJa2009}
Juneja, S., R.~Jain. 2009.
\newblock {The concert/cafeteria queueing problem: a game of arrivals}.
\newblock {\it Proceedings of the Fourth International ICST Conference on
  Performance Evaluation Methodologies and Tools\/}. Fourth International ICST
  Conference on Performance Evaluation Methodologies and Tools, 1--6.

\bibitem[{Juneja and Shimkin(2011)}]{JuSh2011}
Juneja, S., N.~Shimkin. 2011.
\newblock {The Concert Queueing Game with a Finite Homogeneous Population}.
\newblock {\it Operations Research (submitted)\/} .

\bibitem[{Lindsey(2004)}]{Li2004}
Lindsey, R. 2004.
\newblock {Existence, uniqueness, and trip cost function properties of user
  equilibrium in the bottleneck model with multiple user classes}.
\newblock {\it Transportation science\/} {\bf 38}(3) 293.

\bibitem[{Mendelson and Whang(1990)}]{MeWh1990}
Mendelson, H., S.~Whang. 1990.
\newblock {Optimal incentive-compatible priority pricing for the M/M/1 queue}.
\newblock {\it Operations Research\/}  870--883.

\bibitem[{Naor(1969)}]{Na1969}
Naor, P. 1969.
\newblock {The regulation of queue size by levying tolls}.
\newblock {\it Econometrica: journal of the Econometric Society\/} {\bf 37}(1)
  15--24.

\bibitem[{Schmeidler(1973)}]{Sc1973}
Schmeidler, D. 1973.
\newblock {Equilibrium points of nonatomic games}.
\newblock {\it Journal of Statistical Physics\/} {\bf 7}(4) 295--300.

\bibitem[{Whitt(2001)}]{Wh2001}
Whitt, W. 2001.
\newblock {\it {Stochastic Process Limits}\/}.
\newblock Springer.

\end{thebibliography}
\end{document}